\documentclass[11pt]{article}
 \font\bb=msbm10
\usepackage{latexsym}

\usepackage{amssymb}
\usepackage{amsmath}

\begin{document}

\def\cb{\mbox{\bb C}}
\def\tb{\mbox{\bb T}}
\renewcommand{\baselinestretch}{1.2}
\renewcommand{\thefootnote}{\fnsymbol{footnote}}


\renewcommand{\thefootnote}{}

\newcommand{\bt}{\begin{theorem}}
\newcommand{\et}{\end{theorem}}
\newcommand{\bd}{\begin{definition}}
\newcommand{\ed}{\end{definition}}
\newcommand{\bs}{\begin{proposition}}
\newcommand{\es}{\end{proposition}}
\newcommand{\bp}{\begin{proof}}
\newcommand{\ep}{\end{proof}}
\newcommand{\be}{\begin{equation}}
\newcommand{\ee}{\end{equation}}
\newcommand{\ul}{\underline}
\newcommand{\br}{\begin{remark}}
\newcommand{\er}{\end{remark}}
\newcommand{\bex}{\begin{example}}
\newcommand{\eex}{\end{example}}
\newcommand{\bc}{\begin{corollary}}
\newcommand{\ec}{\end{corollary}}
\newcommand{\bl}{\begin{lemma}}
\newcommand{\el}{\end{lemma}}
\newcommand{\bj}{\begin{conjecture}}
\newcommand{\ej}{\end{conjecture}}
\newcommand{\bcex}{\begin{counterexample}}
\newcommand{\ecex}{\end{counterexample}}
\newcommand{\bq}{\begin{question}}
\newcommand{\eq}{\end{question}}

\numberwithin{equation}{section}

\newenvironment{proof}{{\bf Proof. }}{\hfill$\rule{1ex}{1ex}$\par\medskip}

\newcommand{\Ann}{\mbox{Ann}\,}
\newcommand{\coker}{\mbox{Coker}\,}
\newcommand{\Hom}{\mbox{Hom}\,}
\newcommand{\Ext}{\mbox{Ext}\,}
\newcommand{\Tor}{\mbox{Tor}\,}
\newcommand{\Spec}{\mbox{Spec}\,}
\newcommand{\Max}{\mbox{Max}\,}
\newcommand{\Ker}{\mbox{Ker}\,}
\newcommand{\Ass}{\mbox{Ass}\,}
\newcommand{\Assh}{\mbox{Assh}\,}
\newcommand{\Att}{\mbox{Att}\,}
\newcommand{\Supp}{\mbox{Supp}\,}
\newcommand{\MaxSupp}{\mbox{MaxSupp}\,}
\newcommand{\Cosupp}{\mbox{Cosupp}\,}
\newcommand{\gr}{\mbox{grade}\,}
\newcommand{\depth}{\mbox{depth}\,}
\renewcommand{\dim}{\mbox{dim}\,}
\renewcommand{\Im}{\mbox{Im}\,}
\newcommand{\cd}{\mbox{cd}\,}
\newcommand{\q}{\mbox{q}\,}
\newcommand{\amp}{\mbox{amp}\,}
\newcommand{\Min}{\mbox{Min}\,}
\newcommand{\pd}{\mbox{proj.dim}\,}
\newcommand{\id}{\mbox{inj.dim}\,}
\newcommand{\fd}{\mbox{flatdim}\,}
\newcommand{\gd}{\mbox{Gdim}\,}
\newcommand{\h}{\mbox{ht}\,}
\newcommand{\img}{\mbox{img}\,}
\newcommand{\E}{\mbox{E}}
\newcommand{\uhom}{{\mathbf R}\Hom}
\newcommand{\utp}{\otimes^{\mathbf L}}
\newcommand{\ugamma}{{\mathbf R}\Gamma}
\renewcommand{\H}{\mbox{H}}
\newcommand{\V}{\mbox{V}}
\newcommand{\Z}{\mbox{Z}}
\newcommand{\J}{\mbox{J}}
\newcommand{\st}{\stackrel}
\newcommand{\G}{\Gamma}
\newcommand{\D}{\displaystyle}
\newcommand{\U}{\underset}
\newcommand{\R}{\mathbb{R}}
\newcommand{\T}{\mathrm}
\newcommand{\N}{\mathbb{N}}
\newcommand{\lo}{\longrightarrow}
\newcommand{\su}{\subseteq}
\newcommand{\para}{\paragraph}
\newcommand{\ara}{\mbox{ara}}
\newcommand{\fa}{\mathfrak{a}}
\newcommand{\fb}{\mathfrak{b}}
\newcommand{\fI}{\mathfrak{I}}
\newcommand{\fm}{\mathfrak{m}}
\newcommand{\fp}{\mathfrak{p}}
\newcommand{\fq}{\mathfrak{q}}
\newcommand{\fn}{\mathfrak{n}}
\newcommand{\fc}{\mathfrak{c}}
\newcommand{\LH}{Lichtenbaum-Hartshorne Theorem}
\def\mapdown#1{\Big\downarrow\rlap
{$\vcenter{\hbox{$\scriptstyle#1$}}$}}

\newtheorem{theorem}{Theorem}[section]
\newtheorem{acknowledgment}[theorem]{Acknowledgment}
\newtheorem{corollary}[theorem]{Corollary}
\newtheorem{definition}[theorem]{Definition}
\newtheorem{example}[theorem]{Example}
\newtheorem{lemma}[theorem]{Lemma}
\newtheorem{proposition}[theorem]{Proposition}
\newtheorem{remark}[theorem]{Remark}
\newtheorem{conjecture}[theorem]{Conjecture}
\newtheorem{question}[theorem]{Question}


\begin{center}

\def\cl{\centerline}

\cl{\Large{\bf Weighted Gaussian entropy and determinant inequalities}}
\vskip 1 truecm
\cl{\Large{\bf Y.~Suhov$^{1-3}$, S.~Yasaei Sekeh$^{4}$, I. Stuhl$^{5-7}$}}
\vskip .5 truecm

\cl{$^{1}$ DPMMS, University of Cambridge, UK}

\cl{$^{2}$ Math Dept, Penn State University, PA, USA}

\cl{$^{3}$ IPIT RAS, Moscow, RF}

\cl{$^{4}$ DEs, Federal\;University\;of\;S$\tilde{\rm a}$o\;Carlos, SP, Brazil}

\cl{$^{5}$ IMS, University of S\~{a}o Paulo, SP, Brazil}

\cl{$^{6}$ Math Dept, University of Denver, CO, USA}

\cl{$^{7}$ University of Debrecen, Hungary}

\end{center}
\date{\today}

\begin{abstract}
We produce a series of results 
extending  information-theoretical inequalities (discussed by Dembo--Cover--Thomas in 1989-1991) to a weighted version of entropy. The resulting
inequalities involve the Gaussian weighted entropy; they imply a number of new relations
for determinants of positive-definite matrices.
\end{abstract}

\footnotetext{2010 {\em Mathematics Subject Classification:\; \/60A10, 60B05, 60C05.}}
\footnotetext{{\em Key words and phrases:} weight function, weighted entropy, weighted conditional and mutual entropies, weighted Gaussian entropy, weighted determinant inequalities }

\mbox{\quad}

\def\fB{\mathfrak B}
\def\fM{\mathfrak M}
\def\fX{\mathfrak X}
\def\cB{\mathcal B}
\def\cC{\mathcal C}
\def\cM{\mathcal M}
\def\cZ{\mathcal Z}
\def\bu{\mathbf u}
\def\bv{\mathbf v}
\def\bx{\mathbf x}
\def\by{\mathbf y}
\def\om{\omega}
\def\Om{\Omega}
\def\bbP{\mathbb P}
\def\hw{h^{\rm w}}
\def\hwphi{{h^{\rm w}_\phi}}
\def\beq{\begin{eqnarray}}
\def\eeq{\end{eqnarray}}
\def\beqq{\begin{eqnarray*}}
\def\eeqq{\end{eqnarray*}}
\def\rd{{\rm d}}
\def\Dwphi{{D^{\rm w}_\phi}}
\def\Lam{\Lambda}
\def\mwe{{D^{\rm w}_\phi}}
\def\DwPhi{{D^{\rm w}_\Phi}} \def\iw{i^{\rm w}_{\phi}}
\def\bE{\mathbb{E}}
\def\1{{\mathbf 1}}
\def\fB{{\mathfrak B}}
\def\fM{{\mathfrak M}}
\def\diy{\displaystyle}
\def\bbE{{\mathbb E}}
\def\bu{\mathbf u}

\def\BA{{\mathbf A}}
\def\BC{{\mathbf C}}
\def\BI{{\mathbf I}}
\def\BU{{\mathbf U}}
\def\BV{{\mathbf V}}
\def\BX{\mathbf{X}}
\def\uX{{\underline{\BX_{}}}}
\def\BY{{\mathbf Y}}
\def\cY{\mathcal Y}
\def\ux{{\underline{\bx_{}}}}
\def\bPsi{\mathbf{\Psi}}
\def\bsig{\mathbf{\sigma}}
\def\ov{\overline}
\def\oS{{S^{\ct}}}

\def\fB{\mathfrak B}
\def\fM{\mathfrak M}
\def\fX{\mathfrak X}
\def\fY{\mathfrak Y}
\def\fZ{\mathfrak Z}
\def\cB{\mathcal B}
\def\cM{\mathcal M}
\def\cX{\mathcal X}
\def\cY{\mathcal Y}
\def\bu{\mathbf u}
\def\bv{\mathbf v}
\def\bx{\mathbf x}
\def\by{\mathbf y}
\def\bz{\mathbf z}
\def\om{\omega}
\def\Om{\Omega}
\def\bbP{\mathbb P}
\def\bbR{\mathbb R}
\def\hw{h^{\rm w}}
\def\hwphi{{h^{\rm w}_\phi}}
\def\beq{\begin{eqnarray}}
\def\eeq{\end{eqnarray}}
\def\beqq{\begin{eqnarray*}}
\def\eeqq{\end{eqnarray*}}
\def\rd{{\rm d}}
\def\Dwphi{{D^{\rm w}_\phi}}
\def\Lam{\Lambda}
\def\Ups{\Upsilon}
\def\mwe{{D^{\rm w}_\phi}}
\def\DwPhi{{D^{\rm w}_\Phi}}
\def\iw{i^{\rm w}_{\phi}}
\def\bE{\mathbb{E}}
\def\1{{\mathbf 1}}
\def\fB{{\mathfrak B}}
\def\fM{{\mathfrak M}}
\def\diy{\displaystyle}
\def\bbE{{\mathbb E}}
\def\bu{\mathbf u}
\def\lam{\lambda}
\def\bbB{{\mathbb B}}
\def\bbR{{\mathbb R}}
\def\bbS{{\mathbb S}}
\def\bmu{{\mbox{\boldmath${\mu}$}}}

\def\bDel{{\mbox{\boldmath${\Delta}$}}}
\def\bTha{{\mbox{\boldmath${\Theta}$}}}
\def\bPhi{{\mbox{\boldmath${\Phi}$}}}
\def\bPsi{{\mbox{\boldmath${\Psi}$}}}

\def\bbZ{{\mathbb Z}}
\def\fF{\mathfrak F}
\def\bmt{\mathbf t}
\def\B1{\mathbf 1}
\def\b0{\mathbf 0}
\def\beacl{\begin{array}{cl}}
\def\beal{\begin{array}{l}}
\def\beac{\begin{array}{c}}
\def\ena{\end{array}}

\def\uX{{\underline{X_{}}}}
\def\ux{{\underline{x_{}}}}


\def\BA{{\mathbf A}} \def\BB{{\mathbf B}} \def\BC{{\mathbf C}}
\def\BD{{\mathbf D}} \def\BI{{\mathbf I}} \def\BK{{\mathbf K}}
\def\BN{{\mathbf N}} \def\BU{{\mathbf U}} \def\BV{{\mathbf V}}
\def\BX{\mathbf{X}} \def\uX{{\underline{X_{}}}} \def\BY{{\mathbf Y}}
\def\cY{\mathcal Y}  \def\ux{{\underline{x_{}}}}

\def\ct{{\complement}}
\def\bsig{mathbf{\sigma}}\def\ov{\overline} \def\oS{{S^{\ct}}}

\def\BG{\mathbf{G}} \def\BE{\mathbf{E}}\def\Bsigma{\mathbf{\sigma}}
\def\wt{\widetilde} \def\rA{{\rm A}} \def\rT{{\rm T}}

\section{Introduction}

The aim of this paper is to give a number of new bounds involving determinants of positive-definite
matrices. These bounds can be considered as generalizations of
inequalities discussed in \cite{CT, DCT}. A common feature of determinant inequalities (DIs)
from \cite{CT, DCT} is that
most of them have been previously known but often proven by individual arguments (see the bibliography
 in \cite{CT, DCT}). The unifying approach adopted in \cite{CT, DCT} emphasized their common nature
  connected  with/through information-theoretical entropies.

The bounds presented in the current paper are also obtained by a unified method which is based on
{\it weighed entropies} (WEs), more precisely, on Gaussian WEs. Hence, we speak here of weighted
determinant bounds/inequalities. The
weighted determinant inequalities (WDIs) offered in the present paper are novel,
at least to the best of  our knowledge. Moreover, when we choose the weight function to be a (positive)
constant, a WDI become a `standard' DI. In fact, the essence of this work
is that we subsequently examined DIs from \cite{CT, DCT} for a possibility of a (direct) extension
to non-constant weight functions; successful attempts formed the present paper. This reflects
a particular feature of the present paper: a host of new inequalities are obtained by an  old method
while \cite{CT, DCT} re-establish old inequalities by using a new method.

As a primary example, consider the so-called Ky Fan inequality. (We follow the
terminology used in \cite{CT, DCT, CTb}.) This inequality asserts that $\delta (\BC):=\log\,{\rm{det}}\,\BC$
is a concave
function of a positive-definite $d\times d$ matrix $\BC$. In other words, for all strictly positive-definite
$d\times d$ matrices $\BC_1$, $\BC_2$ and $\lam_1,\lam_2\geq 0$ with $\lam_1+\lam_2=1$,
\beq\label{eq:KF} \delta (\lam_1\BC_1+\lam_2\BC_2)-
\lam_1 \delta (\BC_1)-\lam_2\delta (\BC_2)\geq 0;\;\hbox{ equality iff
$\lam_1\lam_2=0$.}\eeq
For original `geometric' proofs of \eqref{eq:KF} and other related inequalities, see Ref \cite{Mo} and the
bibliography therein. In \cite{CT, DCT, CTb} the derivation
of \eqref{eq:KF} occupies few lines and is based on the fact that under a variance constraint, the
differential entropy is maximized at a Gaussian density.

A weighted Ky Fan inequality \eqref{eq:wKF}\ has been proposed in \cite{SY}, Theorem 3.2;
the derivation is
also short and based on a maximization property of the weighted entropy (cf.
Theorem \ref{Lemma 2.1} below). Namely, given $\BC_1$, $\BC_2$ and $\lam_1,\lam_2$ as above and a nonnegative function
$\bx\in\bbR^d\mapsto\phi (\bx)$, positive on an open domain in $\bbR^d$, assume condition
 \eqref{eq:WKFCond}. Then
\beq\label{eq:wKF}\sigma (\lam_1\BC_1+\lam_2\BC_2)-\lam_1\sigma (\BC_1 )
-\lam_2\sigma (\BC_2 )
\geq 0;\;\hbox{ equality again iff $\lam_1\lam_2=0$.}\eeq
Here, for a strictly positive-definite $\BC$, the value $\sigma (\BC)=\sigma_\phi (\BC)$ is as follows:
\beq\label{eq:sigm}
\sigma_\phi (\BC)=\frac{\alpha_\phi (\BC)}{2} \log \left[ (2\pi)^d ({\rm{det}}\,\BC)\right]
+\frac{\log\,e}{2}{\rm{tr}}\,\BC^{-1}\bPhi_{\BC ,\phi}:=\hwphi (f^{\rm{No}}_{\BC}).\eeq
Next, $\alpha_\phi (\BC)>0$ and positive-definite
matrix $\bPhi_{\BC ,\phi}$ are given by
\beq\label{eq:bPhi}\alpha_\phi (\BC)=\int\limits_{\bbR^d}\phi (\bx_1^d)
f^{\rm{No}}_{\BC}(\bx_1^d)\rd\bx_1^d,\;\; \bPhi_{\BC ,\phi}
=\int\limits_{\bbR^d}\bx_1^d\,\left(\bx_1^d\right)^{\rm T}\phi (\bx_1^d)
f^{\rm{No}}_{\BC}(\bx_1^d)\rd\bx_1^d,\eeq
and $f^{\rm{No}}_{\BC}$ stands for a normal probability density function (PDF) with mean $\b0$ and
covariance matrix $\BC$:
\beq\label{eq:fBCii} f^{\rm{No}}_\BC (\bx)=\frac{1}{(2\pi )^{d/2}\big({\rm{det}\,\BC\big)^{1/2}}}
\exp\,\left(-\frac{1}{2}\,{\bx}^{\rm T}\BC^{-1}\bx\right),\;\;\bx =\begin{pmatrix}x_1\\
\vdots\\ x_n\end{pmatrix} \in\bbR^d.\eeq
In terms of a multivariate normal random vector $\BX_1^d\sim f^{\rm{No}}_\BC$: $\alpha_\phi (\BC)=
\bbE\phi (\BX_1^d)$ and $\bPhi_{\phi ,\BC}= \bbE\left(\phi (\BX_1^d)\left[\BX_1^d\left(\BX_1^d
\right)^{\rm T}\right]\right)$.
In \eqref{eq:fBCii} and below we routinely omit the indices in the notation like $\bx_1^d$ and
$\BX_1^d$. The quantity $\hwphi (f^{\rm{No}}_{\BC})=-\int\limits_{\bbR^d}\phi (\bx )f^{\rm{No}}_{\BC}(\bx )
\log\,f^{\rm{No}}_{\BC}(\bx )\rd\bx$ is the
{\it weighted entropy} of $f^{\rm{No}}_{\BC}$ with weight function $\phi$, a concept analyzed in
detail below. For $\phi (\bx )\equiv 1$, $\hwphi (f^{\rm{No}}_{\BC})$ coincides with a `standard'
(differential) entropy of a normal PDF.

The assumption upon $\BC_1$, $\BC_2$ and $\lam_1$, $\lam_2$ consists of two bounds and reads
\beq\label{eq:WKFCond}\beac
\diy\lam_1\alpha (\BC_1)+\lam_2\alpha (\BC_2)-\alpha (\lam_1\BC_1+\lam_2\BC_2 )\geq 0,\\
\beal\diy\Big[\lam_1\alpha (\BC_1)+\lam_2\alpha (\BC_2)-\alpha (\lam_1\BC_1+\lam_2\BC_2 )\Big]\\
\qquad\times\log \Big\{ (2\pi)^d\Big[{\rm{det}}\, (\lam_1\BC_1+\lam_2\BC_2 )\Big]\Big\}
 +{\rm{tr}}\,\Big[(\lam_1\BC_1+\lam_2\BC_2 )^{-1}\bDel\Big]\leq 0\ena\ena \eeq
where matrix $\bDel=\lam_1\bPhi_{\BC_1}+\lam_2\bPhi_{\BC_2}
-\bPhi_{\lam_1\BC_1+\lam_2\BC_2}$. Bounds \eqref{eq:WKFCond} have opposite directions
and stem from the weighted Gibbs inequality. Cf.  Eqns (1.3),
(3.3) from Ref \cite{SY} and \eqref{3.3}, \eqref{eq:Cnd1Gbbs} from Section \ref{Sect:GWE} below.

When  $\phi (\bx )\equiv 1$, Eqn \eqref{eq:WKFCond} is satisfied: we have equalities.
In this case the weighted Ky Fan inequality \eqref{eq:wKF}
transforms into \eqref{eq:KF}. In general, condition \eqref{eq:WKFCond} is not trivial: in a simplified
case of an
exponential weight function $\phi (\bx )=
\exp\left(\bmt^{\rm T}\bx\right)$, $\bmt\in\bbR^d$, it has been analyzed, both analytically
and numerically, in \cite{SY1}. (Here, $\phi (\bx )\equiv 1$ means $\bmt =\b0$.) As was shown in
\cite{SY1}, for given $\BC_1$, $\BC_2$, $\lam_1$, $\lam_2$ and
$\phi$ (that is, for a given $\bmt$), Eqn \eqref{eq:WKFCond} may or may not be fulfilled. (And when
\eqref{eq:WKFCond} fails, \eqref{eq:wKF} may still hold true.) Moreover, when
\eqref{eq:WKFCond}  holds, it may or may not produce a strictly positive expression in the RHS of
bound \eqref{eq:KF}. (Thus, in some cases  we can speak of an improvement in the Ky Fan inequality.)
See Ref \cite{SY1}.
We believe that further studies in this direction should follow, focusing on specific forms of weight
function $\phi$.

In our opinion, this paper paves way to a similar analysis of the whole host of newly
established WDIs. These inequalities should be taken with a justified degree of caution: offered
sufficient conditions (stated in the form of bounds involving various weight function) may fail for
particular $\BC_1$, $\BC_2$, $\lam_1$, $\lam_2$, and $\phi$, and a given WDI may
or may not yield an improvement compared to its `standard' counterpart. For reader's convenience
we list the sufficient conditions figuring across the paper: Eqns \eqref{eq:condSi}, \eqref{cond.MWE},
\eqref{3.3}, \eqref{eq:Cnd1Gbbs},
\eqref{eq:1.10}, \eqref{eq:1.16}  \eqref{eq:1.161}, \eqref{eq:sumcond},
\eqref{eq:sumcondi}, \eqref{eq:cndGSiN}, \eqref{eq:Hada}, \eqref{cond.MWEN},
\eqref{eq:ExtHAss}, \eqref{eq:CndCncv}, \eqref{eq:CndCncv2}  and \eqref{eq:condvarpi}.

The presented WDIs generalize what is sometimes called elementary information-theoretic
inequalities. An opposite example is the entropy-power inequality; and related bounds.
Here the intuition is more intricate; some initial results have been proposed in
\cite{SYK}.

The paper is organized as follows. In Section \ref{sect:WERSRWF} we work with a general setting,
elaborating on properties of weighted entropies which have been established earlier
in \cite{SY}. Section \ref{Sect:GWE} summarizes
some properties of Gaussian weighted entropies while Section \ref{sect:WEM} analyzes the behavior
of weighted entropies under mappings; these sections also rely on Ref. \cite{SY}. The WDIs
are presented in Sections \ref{sect:MWDIs} and \ref{sect:WHI} as a sequel to the material from
Section \ref{sect:WERSRWF} --  \ref{sect:WEM}. Again, for reader's convenience we list them
here as Eqns \eqref{eq2:1}, \eqref{eq:m}, 
\eqref{eq:s}, \eqref{eq:w}, \eqref{eq:u}, \eqref{eq:z}, \eqref{eq:WSHI}, \eqref{eq;mu} and \eqref{eq:varpiAB}.

\section{Random strings and reduced weight functions}\label{sect:WERSRWF}

The WE of a probability distribution was introduced in late 1960s -- early 1970s;
see, e.g., \cite{BG}. (Another term that can be used
is a context-dependent or a preferential entropy.)
The reader is referred to  \cite{SY} where a number of notions and elementary inequalities were
established for the WE, mirroring well-known facts about the standard (Shannon) entropy. We also use Refs \cite{CT, DCT}
as a source of standard inequalities which we extend to the
case of the WE. To keep pre-emptiveness, we follow the system of notation from \cite{CT, DCT, SY}
with minor deviations.

Let us begin with general definitions.
The WE of a random element $X$ taking values in a standard measure space (SMS) $(\cX,\fM,\nu)$ with a weight function (WF) \ $x\in\cX\mapsto\phi (x)\geq 0$ is defined by
\beq\label{eq:WEdef}\hwphi (X)=\hwphi (f)=\bbE \left(\phi (X)\log\,f(X)\right)=-\int\limits_\cX\phi (x)f(x)\log\,f(x)\nu (\rd x),
\eeq
assuming that $\phi$ is measurable and the integral is absolutely convergent. Here $f=f_X$ is the probability mass/density function
(PM/DF) of $X$ relative to measure $\nu$. Symbol $\bbE$ stands for the expected value (relative to a
probability distribution that is explicitly specified or emerges from the context in an unambiguous
manner).

A number of properties of the WE are related to a Cartesian product structure.
Let random elements $X_1,\dots, X_n$ be given, taking values in SMSs
$(\cX_i,\fM_i,\nu_i)$, $1\leq i\leq n$. Set $\uX_1^n:=\{X_1,\dots, X_n\}$ and assume that
$X_1,\dots, X_n$ have a joint PM/DF $f_{\uX_1^n}(\ux_1^n)$, \ $\ux_1^n\in
\cX_1^n:=\operatornamewithlimits{\times}\limits_{1\leq i\leq n}\cX_i$, relative to the measure
$\nu_1^n:=\operatornamewithlimits{\times}\limits_{1\leq i\leq n}\nu_i$; for brevity we will sometimes
set $f_{\uX_1^n}=f$. The joint WE of string $\;\uX_1^n\;$ is defined as
\beq\label{eq:jntWEdef}\hwphi (\uX_1^n )=-\bbE \left( \phi (\uX_1^n)\log\,f(\uX_1^n) \right)=-\int\limits_{\cX_1^n}\phi (\ux_1^n)f(\ux_1^n)\log\,f(\ux_1^n)
\nu_1^n(\rd\ux_1^n).\eeq
Given a set $S\subseteq I:=\{1,2,\dots,n\}$, write
\beq\label{eq:Sc1}\hbox{$\uX(S)$, $\uX (\oS)$ for strings $\{X_i:i\in S\}$,
$\{X_i:i\in\oS\}$, respectively, where $\oS=I\setminus S$.}\eeq
Next,  let $\ux (S)$ and $\ux (\oS )$ stand for
\beq\label{eq:Sc2}\hbox{$\{x_i:\;i\in S\}\in\cX(S):=\operatornamewithlimits{\times}\limits_{i\in S}\cX_i$ and
$\{x_i:\;i\in\oS\}\in\cX(\oS):=\operatornamewithlimits{\times}\limits_{i\in\oS}\cX_i$.}\eeq
Accordingly, the marginal
PD/MF $f_{\uX (S)}(\ux (S))$ emerges, for which we will often write $f_S(\ux (S))$ or even $f(\ux (S))$
for short. Furthermore, given a WF $\ux_1^n\mapsto\phi (\ux_1^n)\geq 0$, we define the function
$\;\psi (S):\ux (S)\mapsto\psi (S;\ux (S))\geq 0$ involving
the conditional PM/DF $f_{\uX (\oS)|\uX (S)}\Big(\ux(\oS)\,|\,\ux(S)\Big)$:
\beq\label{eq:psiS} \psi (S;\ux (S))=\int\limits_{\cX (\oS)} \phi (\ux_1^n)f_{\uX (\oS)|\uX (S)}
\Big(\ux(\oS)\,|\,\ux(S)\Big)\nu_{\cX (\oS)}(\rd\ux(\oS))\eeq
where $\nu_{\cX (\oS)}:=\operatornamewithlimits{\times}\limits_{i\in\oS}\nu_i$. For brevity we again write
sometimes $f_{\oS|S}$\; instead of $f_{\uX (\oS )|\uX (S)}$ or omit subscripts altogether. We also write
$\rd\ux(S)$ and $\rd\ux(\oS)$ instead of $\nu_{\cX (S)}(\rd\ux(S))$ and $\nu_{\cX (S)}(\rd\ux(\oS))$ and
$\rd\ux$ instead of $\nu_1^n(\rd\ux_1^n)$.

Function $\psi (S;\,\cdot\,)$ will play the role of a reduced (or induced) WF when we pass from
$\uX_1^n$ to a sub-string $\uX (S)$. More precisely, set
\beq\label{eq:jntWEsubc}\beacl\hw_{\psi (S)} (\uX (S))&=-\bbE \left( \psi (S;\uX (S)\log f_S(S;\uX (S)) \right)\\
\;&\diy=-\int\limits_{\cX (S)}\psi (S;\ux (S))f_S(\ux (S))\log\,
f_S(\ux (S))\rd\ux (S),\ena\eeq
with $\nu_{\cX (S)}:=\operatornamewithlimits{\times}\limits_{i\in S}\nu_i$. Cf. \cite{SY}.
 Next, for $k=1,\ldots ,n$ define
\beq \label{averageWE} h^{{\rm w},n}_k=\diy\binom{n}{k}^{-1} \diy\sum\limits_{S\subseteq I:\;\# (S)=k}
\frac{h^{\rm w}_{\psi(S)}(\uX(S))}{k}.\eeq
(Here and below, $\# (S)$  and $\# (\oS)$ are the cardinalities of $S$ and $\oS$.)
Here $h^{{\rm w},n}_k$ renders the averaged WE (per string and per element) of a randomly drawn
$k$-element sub-string in $\uX_1^n$.

In what follows we use the concepts of the conditional and mutual WE and their properties; cf.
\cite{SY}. These objects are used with a host of WFs, depending on the context. Consider the following
 condition:
\beq\label{eq:condSi}\beal\hbox{$\forall$ $i\in S\subseteq I$,
with $S^-_i=\{j\in S:\;j<i\}$ and $S^+_i=\{j\in S:\;j>i\}$,}\\
\quad\diy\int\limits_{\cX (S)}\psi (S;\ux (S))\Big\{f(\ux (S))-f(\ux (S^-_i))\times\Big[f(x_i|\ux (S^-_i))
f(\ux (S^+_i)|\ux (S^-_i))
\Big]\Big\}\rd\ux (S)\geq 0,\ena\eeq
with standard agreements when one of the sets $S^\pm_i=\emptyset$. Pictorially, Eqn \eqref{eq:condSi}
is an extension of bound (1.27) from \cite{SY}; it means
that for all $i\in S\subseteq I$, the induced WF $\psi (S;\,\cdot\,)$ is correlated more positively with the
marginal PM/DF $f_S(\ux (S))$ than with the dependence-broken product
$f_{S^-_i}(\ux (S^-_i))\times\Big[f_{\{i\}|S^-_i}(x_i|\ux (S^-_i))f_{S^+_i|S^-_i}(\ux (S^+_i)|\ux (S^-_i))\Big]$.
Another version of (essentially) the same property is Eqn \eqref{cond.MWE} below.

\br {\rm The special choice of sets $S^\pm_i$ is not particularly important: it can be a general
partition
of $S\setminus\{i\}$ allowing us to use the chain rule for the conditional WE (see below).}
\er


\bt \label{average.WE}
{\rm{(Cf. \cite{CT}, Lemma 7 or \cite{DCT}, Theorem 1.)}} Let $h^{{\rm w},n}_k$ be defined as in {\rm{(\ref{averageWE})}} and
assume \eqref{eq:condSi}. Then
\beq h^{{\rm w},n}_1 \geq h^{{\rm w},n}_2\geq \ldots \geq h^{{\rm w},n}_{n-1} \geq h^{{\rm w},n}_n.\eeq
\et

\bp Begin with the last inequality, $h^{{\rm w},n}_{n-1} \geq h^{{\rm w},n}_n$. Let
$1\leq i\leq n$ and choose $S=I$, $S^-_i=I^-_i:=\{1,\ldots ,i-1\}$ and $S^+_i=I^+_i:=\{i+1,\ldots ,n\}$, with
$\{i\}^{\ct}=I^-_i\cup I^+_i$ (cf. \eqref{eq:Sc1}, \eqref{eq:Sc2}). Then the condition
$$\int\limits_{(\cX^n)^1}\phi (\ux)\left[f (\ux )-f_{\uX_1^{i-1}}(\ux_{\,1}^{i-1})f(x_i|\ux_{\,1}^{i-1})f(\ux_{\,i+1}^n|\ux_{\,1}^{i-1})\right]\rd\ux\geq 0\;\hbox{ (by virtue of \eqref{eq:condSi}),}$$
yields:
\beqq\hwphi(\uX_1^n)&=\;\hwphi (X_i|\uX (\{i\}^{\ct})+h^{\rm w}_{\psi (\{i\}^{\ct})}(\uX (\{i\}^{\ct}))&\hbox{ by the chain rule}\\
&\;\leq\; h^{\rm w}_{\psi (I^-_i)}(X_i|\uX_1^{i-1})+h^{\rm w}_{\psi (\{i\}^{\ct})}(\uX (\{i\}^{\ct}))&
\;\hbox{by Lemma 1.3 from \cite{SY}.}\eeqq
Here reduced WFs $\psi (\{i\}^{\ct})$ and $\psi (I^-_i)$ are calculated according to the recipies
in \eqref{eq:psiS}, \eqref{eq:jntWEsubc}.


Taking the sum, we obtain:
\beq\label{Inq:2.1} n\,\hwphi(\uX_1^n)\leq \sum\limits_{i=1}^n h^{\rm w}_{\psi (\{i\}^{\ct})}(\uX (\{i\}^{\ct})+\sum\limits_{i=1}^n h^{\rm w}_{\psi (I^-_i)}(X_i|\uX_1^{i-1}).\eeq
By using the chain rule, $\;\diy\sum\limits_{i=1}^n h^{\rm w}_{\psi (I^-_i)}(X_i|\uX_1^{i-1})=\hwphi (\uX_1^n)$.
Hence, Eqn \eqref{Inq:2.1} becomes
\beqq (n-1) \hwphi(\uX_1^n)\leq\sum\limits_{i=1}^n h^{\rm w}_{\psi (\{i\}^{\ct})}(\uX (\{i\}^{\ct}).\eeqq
Consequently,
\beq\label{Ine2:2} \hwphi(\uX_1^n) \leq\sum\limits_{i=1}^n\frac{h^{\rm w}_{\psi (\{i\}^{\ct})}(\uX (\{i\}^{\ct})}{n-1},\eeq
which yields that $h^{{\rm w},n-1}_{n-1} \geq h^{{\rm w},n}_n$.

This argument can be repeated if we restrict the WE and the PM/DF to a $k$-element subset $S=\{i_1,\ldots ,i_k\}\subset I$ listed in an increasing order of its points and perform a uniform choice over its $(k-1)$-elements subsets.
Condition \eqref{eq:condSi}
yields the bound
\beqq\frac{1}{k}\hw_{\psi (S)}(\uX (S))\leq\frac{1}{k}\sum_{i\in S}\frac{\hw_{\psi (S\setminus\{i\})}(\uX (S\setminus\{i\}))}{k-1}.\eeqq

Hence for each $k$-element subset, $h^{{\rm w},k}_{k-1} \geq h^{{\rm w},k}_k$. Therefore, the inequality remains true after taking the average over all $k$-element subsets drawn uniformly. \ep

In Theorem \ref{lemma.g} we extend the result of Theorem \ref{average.WE} to exponents
of WEs for sub-strings in $\uX_1^n$.

\bt\label{lemma.g}
{\rm{(Cf. \cite{CT}, Corollary of Lemma 7 or \cite{DCT}, Corollary 1)}} Given $r>0$, define:
\beq g^{{\rm w},n}_k=\binom{n}{k}^{-1} \diy\sum\limits_{S\subseteq I:\;\# (S)=k}\exp\,\left[{r\;\frac{h^{\rm w}_{\psi(S)}(\uX(S))}{k}}\right].\eeq
Then, under assumption \eqref{eq:condSi},
\beq\label{eq:gnn}g^{{\rm w},n}_1 \geq g^{{\rm w},n}_2\geq \dots \geq g^{{\rm w},n}_{n-1} \geq
g^{{\rm w},n}_n.\eeq
\et

\bp Again, it is convenient to  start with the last bound in \eqref{eq:gnn}. As in \cite{CT}, multiply Eqn (\ref{Ine2:2}) by $r$, exponentiate and apply the arithmetic--geometric mean inequality to obtain $g^{{\rm w},n}_{n-1} \geq g^{{\rm w},n}_n$. The result is then completed with the help of same argument as in the proof of Theorem \ref{average.WE}.
\ep

In Theorem \ref{average.CWE} we analyse  the averaged conditional WEs
for sub-strings in $\uX_1^n$.

\bt\label{average.CWE}
{\rm{(Cf. \cite{DCT}, Theorem 2.)}} Let $p^{{\rm w},n}_k$ be defined as
\beq \label{averageCWE} p^{{\rm w},n}_k=\binom{n}{k}^{-1} \diy\sum\limits_{S\subseteq I:\;\# (S)=k}
\frac{h^{\rm w}_{\phi}(\uX(S)|\uX(\oS))}{k}.\eeq
Then under the assumption
\beq\label{condGR} \int\limits_{\cX_1^n} \phi(\ux)\left[ f(\ux)-\prod\limits_{i=1}^n f(x_i)\right]\rd\ux \geq 0\eeq
we have that
\beq p^{{\rm w},n}_1 \leq p^{{\rm w},n}_2\leq \ldots \leq p^{{\rm w},n}_{n-1} \leq p^{{\rm w},n}_n.\eeq
\et

\bp Following the argument used in \cite{SY}, Theorem 3.1, condition (\ref{condGR}) yields
\beqq \hwphi (\uX_1^n )\leq \sum\limits_{i=1}^n h^{\rm w}_{\psi(\{i\})}(X_i).\eeqq
Subtracting both sides from $n \hwphi (\uX_1^n )$, we obtain:
\beqq (n-1)\hwphi (\uX_1^n )\geq \sum\limits_{i=1}^n \left[\hwphi (\uX_1^n )- h^{\rm w}_{\psi(\{i\})}(X_i)
\right],\eeqq
By the conditional WE definition,
\beqq \hwphi (\uX_1^n )=\hwphi (\uX_1^{i-1},\uX_{i+1}^n|X_i)+ h^{\rm w}_{\psi(\{i\})}(X_i).\eeqq
Hence,
\beq\label{Ieq S.1} (n-1)\hwphi (\uX_1^n )\geq \sum\limits_{i=1}^n \hwphi (\uX_1^{i-1},\uX_{i+1}^n|X_i).\eeq
Dividing (\ref{Ieq S.1}) by $n(n-1)$ yields that $p^{{\rm w},n}_{n-1} \leq p^{{\rm w},n}_n$. Finally,
applying the same argument as in Theorem \ref{average.WE} completes the proof.
\ep

The next step is to pass to mutual WEs.

\bt {\rm{(Cf. \cite{DCT}, Corollary 2.)}} Consider the averaged mutual WE between a subset (or a
sub-string) and its complement:
\beq \label{averagemutualWE} q^{{\rm w},n}_k=\binom{n}{k}^{-1} \diy\sum\limits_{S\subseteq I:\;\# (S)=k}
\frac{i^{\rm w}_{\phi}\Big(\uX(S):\uX(\oS)\Big)}{k},\eeq
and assume \eqref{eq:condSi}. Then
\beq q^{{\rm w},n}_1 \geq q^{{\rm w},n}_2\geq \ldots \geq q^{{\rm w},n}_{n-1} \geq q^{{\rm w},n}_n.\eeq
\et

\bp The result is straightforward, from Theorems \ref{average.WE}
and \ref{average.CWE} and the following relation between conditional and mutual WEs:
\beqq i^{\rm w}_{\phi}\Big(\uX(S):\uX(\oS)\Big)=h^{\rm w}_{\psi(S)}(\uX(S))-\hwphi
\Big(\uX(S)\,\big|\,\uX(\oS)\Big).\eeqq
\ep

In Theorem \ref{averagemutualWE2} we consider the following condition:
for all set $S$ with $\#\,S\geq 2$ and $i,j\in S$ with $i\neq j$,
\beq\label{cond.MWE}\beal\diy\int\limits_{\cX_1^n}\phi(\ux) f(\ux(\oS)|\ux (S))
\Big[f(\ux(S))\\
\qquad\quad -f(\ux (S\setminus\{i,j\}))\;f(x_i|\ux (S\setminus\{i,j\}))\;f(x_j|\ux (S\setminus\{i,j\}))\Big]
\rd\ux \geq 0.\ena\eeq
The meaning of \eqref{cond.MWE} is that for all $S$ and $i,j$ as above, the reduced WF
$\psi_S(\ux (S))$ is correlated more positively with $f(\ux(S))$ than with the PM/DF
$f(\ux (S\setminus\{i,j\}))\;f(x_i|\ux (S\setminus\{i,j\}))\;f(x_j|\ux (S\setminus\{i,j\}))$
where the conditional
dependence between $X_i$ and $X_j$ is broken, given $\uX (S\setminus\{i,j\})$.

\bt\label{averagemutualWE2}
{\rm{(Cf. \cite{DCT}, Theorem 3.)}} Define the average mutual WE as
\beq\label{average MWE} I^{{\rm w},n}_k =\binom{n}{k}^{-1} \diy\sum\limits_{S\subseteq I:\;\# (S)=k}
i^{\rm w}_{\phi}\Big(\uX(S):\uX(\oS)\Big).\eeq
By symmetry of the mutual WE, $I^{{\rm w},n}_k=I^{{\rm w},n}_{n-k}$. Assume condition \eqref{cond.MWE}.
Then
\beq I^{{\rm w},n}_1 \leq I^{{\rm w},n}_2\leq \ldots \leq
I^{{\rm w},n}_{\lfloor n/2\rfloor}.\eeq
\et

\bp Let $k\leq \lfloor n/2\rfloor$. If $S$ is a subset of size $k$ then $S$ has $k$ subsets of size $k-1$. Thus, we write:
 \beqq \begin{array}{l}\diy k\;i^{\rm w}_{\phi}\Big[\uX(S):\uX(\oS)\Big]-\sum\limits_{j\in S}\; i^{\rm w}_{\phi}\Big[\uX(S_j):\uX((S_j)^{\ct})\Big]\\
 \quad\quad =\diy \sum\limits_{j\in S}\left\{ i^{\rm w}_{\phi}\Big[(\uX(S_j),X_j):\uX(\oS)\Big]-i^{\rm w}_{\phi}\Big[\uX(S_j):(\uX(\oS),X_j)\Big]\right\}. \end{array} \eeqq
After direct computations, we obtain:
\beqq i^{\rm w}_{\phi}\Big[(\uX(S_j),X_j):\uX(\oS)\Big]= i^{\rm w}_{\psi(S_j \cup\oS)}
\Big[\uX(S_j):\uX(\oS)\Big]
+i^{\rm w}_{\phi}\Big[X_j:\uX(\oS)|\uX(S_j)\Big],\eeqq
and
\beqq i^{\rm w}_{\phi}\Big[\uX(S_j):(\uX(\oS),X_j)\Big]=i^{\rm w}_{\psi(S_j \cup \oS)}\Big[\uX(S_j):\uX(\oS)\Big]
+i^{\rm w}_{\phi}\Big[X_j:\uX(S_j)|\uX(\oS)\Big].\eeqq
Here $i^{\rm w}_{\phi}\Big[X_j:\uX(\oS)|\uX(S_j)\Big]$, $i^{\rm w}_{\phi}\Big[X_j:\uX(S_j)
|\uX(\oS)\Big]$ are mutual-conditional WEs emerging as in the proof of Theorem 3 from \cite{DCT}:
\beq\label{eq:psistar1}\beacl
i^{\rm w}_{\phi}\Big[X_j:\uX(\oS)|\uX(S_j)\Big]&\diy=\bbE \Big(\phi (\uX)\log\frac{f(X_j,\uX (\oS)|\uX (S_j))}{f(X_j|\uX (S_j))f(\uX(\oS)|\uX (S_j))}\Big)\\ \;&\;\\
\;&=\diy\int\limits_{\cX_1^n}\phi(\ux)f(\ux) \log\;\frac{f(x_j,\ux(\oS)|\ux(S_j))}{f(x_j|\ux(S_j)) f(\ux(\oS)|\ux(S_j))}\;\rd\ux,\ena\eeq
\beq\label{eq:psistar2}\beacl
i^{\rm w}_{\phi}\Big[X_j:\uX(S_j)|\uX(\oS)\Big]&\diy=\bbE \Big(\phi (\uX ) \log\;\frac{f(X_j,\uX (S_j)|\uX (\oS))}{f(X_j|\uX (\oS)) f(\uX (S_j)|\uX (\oS))}\Big)\\ \;&\;\\
\;&=\diy\int\limits_{\cX_1^n}\phi(\ux)f(\ux) \log\;\frac{f(x_j,\ux(S_j)|\ux(\oS))}{f(x_j|\ux(\oS)) f(\ux(S_j)|\ux(\oS))}\;\rd\ux.
\ena\eeq

In the remaining argument we will make an extensive use of definition \eqref{eq:psiS}, employing
WF $\psi (S)$ for a number of choices of set $S$.

Using mutual-conditional WEs we can write:
 \beq \begin{array}{l} k\;i^{\rm w}_{\phi}\Big[\uX(S):\uX(\oS)\Big]-\diy\sum\limits_{j\in S}\;i^{\rm w}_{\phi}
 \Big[\uX(S_j):\uX((S_j)^{\ct})\Big]\\
\quad\quad =\diy \sum\limits_{j\in S}\left\{i^{\rm w}_{\phi}\Big[X_j:\uX(\oS)|\uX(S_j)\Big]-i^{\rm w}_{\phi}\Big[X_j:\uX(S_j)|\uX(\oS)\Big]\right\}\\
\quad\quad = \diy \sum\limits_{j\in S} \Big[ h^{\rm w}_{\psi(S)}(X_j|\uX(S_j))-\hwphi(X_j|\uX(\oS),
\uX(S_j))\\
\quad\quad\qquad\qquad -\diy h^{\rm w}_{\psi(j\cup\oS)}(X_j|\uX(\oS))-\hwphi(X_j|\uX(\oS),\uX(S_j))\Big]\\
\quad\quad = \diy \sum\limits_{j\in S} \left[h^{\rm w}_{\psi(S)}(X_j|\uX(S_j))-h^{\rm w}_{\psi(j\cup\oS)}
(X_j|\ux(\oS))\right].\end{array}\eeq
Summing over all subsets of size $k$ and reversing the order of summation, we obtain:
\beq\label{Eq S.2}\begin{array}{l}\diy\sum\limits_{S\subseteq I:\;\# (S)=k}\left\{k\;i^{\rm w}_{\phi}\Big[\uX(S):
\uX(\oS)\Big]-\diy\sum\limits_{j\in S} i^{\rm w}_{\phi}\Big[\uX(S_j):\uX((S_j)^{\ct})\Big]\right\}\\
\qquad\qquad =\diy \sum\limits_{j=1}^n \diy\sum\limits_{S\subseteq I:\;\# (S)=k, j\in S} \left[[h^{\rm w}
_{\psi(j\cup S_j)}(X_j|\uX(S_j))-h^{\rm w}_{\psi(j\cup\oS)}(X_j|\uX(\oS))\right].\end{array}\eeq
The RHS of (\ref{Eq S.2}) can be rewritten in the following way:
\beqq \diy\sum\limits_{j=1}^n \diy\sum\limits_{S':\;\# (S')=k-1, j\not\in S}\left[h^{\rm w}_{\psi(S'\cup j)}
(X_j|\uX(S'))-h^{\rm w}_{\psi(\{S^\prime\cup j\}^{\ct} \cup j)}(X_j|\uX(\{S'\cup j\}^{\ct}))\right],\eeqq
or equivalently
\beqq\begin{array}{l} \diy\sum\limits_{j=1}^n \left[\diy\sum\limits_{S':\;\# (S')=k-1, S'\subset\{j\}^{\ct}}
h^{\rm w}
_{\psi(S'\cup j)}(X_j|\uX(SÕ))\right.
\left.-\diy\sum\limits_{{S^{\prime\prime}}:\;\# ({S^{\prime\prime}})=n-k, {S^{\prime\prime}}\subset\{j\}^{\ct}} h^{\rm w}_{\psi({S^{\prime\prime}}\cup j)}(X_j|
\uX({S^{\prime\prime}}))\right].\end{array}\eeqq

Since $k\leq \lfloor n/2\rfloor$, then $k-1<n-k$. A set $S^{\prime\prime}$ with $n-k$ elements has $\diy
\binom{n-1}{k-1}$ subsets of size $k-1$. Owing to Lemma 1.3 from \cite{SY}, for each such subset
$\wt S\subset S^{\prime\prime}$, under
assumption (\ref{cond.MWE}) we have that
\beq\label{eq:Ucond.MWE} h^{\rm w}_{\psi({S^{\prime\prime}}\cup j)}(X_j|\uX({S^{\prime\prime}}))
 \leq  h^{\rm w}_{\psi({\wt S}\cup j)}(X_j|\uX({\wt S})).\eeq
With the same argument as in \cite{DCT} we conclude from \eqref{eq:Ucond.MWE} that
\beqq \diy\sum\limits_{S\subseteq I:\;\# (S)=k}\left\{k\;i^{\rm w}_{\phi}\Big[\uX(S):\uX(\oS)\Big]-\diy
\sum\limits_{j\in S} i^{\rm w}_{\phi}\Big[\uX(S_j):\uX((S_j)^{\ct})\Big]\right\} \geq 0.\eeqq
Then, since each set of size $k$ occurs $n-k+1$ times in the second sum, we can write
\beqq \diy k\sum\limits_{S\subseteq I:\;\# (S)=k} i^{\rm w}_{\phi}(\uX(S):\uX(\oS))\geq (n-k+1)\diy
\sum\limits_{S'\subseteq I:\;\# (S')=k-1}i^{\rm w}_{\phi}(\uX(S'):\uX(S'^{\ct})).\eeqq
Dividing by $\diy k \binom{n}{k}$ concludes the proof.
\ep

\section{Gaussian weighted entropies }\label{Sect:GWE}

As we said in the introduction, the WDIs are connected with the Gaussian WE $\hwphi (f^{\rm{No}}_{\BC}):=
-\int\limits_{\bbR^d}\phi (\bx )f^{\rm{No}}_{\BC}(\bx )
\log\,f^{\rm{No}}_{\BC}(\bx ) \rd\bx$; cf. \eqref{eq:sigm}, \eqref{eq:fBCii}.
Throughout the paper we use a number of properties established in \cite{SY}.
One of them is maximization of the WE
$\hwphi (f):=-\int\limits_{\bbR^d}\phi (\bx )f(\bx )\log\,f(\bx )\rd\bx$
at $f=f^{\rm{No}}_\BC$. More precisely, consider the following inequalities
\beq\label{3.3}\begin{array}{c}\diy\int_{\bbR^d}\phi (\bx )\Big[f(\bx )-f^{\rm{No}}_\BC (\bx )\Big]\rd\bx\geq 0\\
\diy\log \left[ (2\pi)^d ({\rm{det}}\,\BC)\right]\int_{\bbR^d}\phi (\bx )\Big[f(\bx )-f^{\rm{No}}_\BC (\bx )\Big]\rd
\bx +{\rm{tr}}\,\Big[\BC^{-1}\left(\bPhi^{\rm{No}}_\BC-\bPhi\right)\Big]\leq 0.
\end{array}\eeq

\bt\label{Lemma 2.1} Let $\BX=\BX_1^d\sim f(\bx)$, $\bx\in\bbR^d$, be a random vector with
PDF $f$, mean zero and covariance matrix
$$\BC=\bbE_\BC\Big(\big(\BX_1^d\big)\,\big({\BX_1^d})^{\rm T}\Big) =
\diy\int\limits_{\bbR^d}\bx {\bx }^{\rm T}f^{\rm{No}}_{\BC}(\bx )
\rd\bx .$$
Set:
$$\diy\bPhi =\bbE_\BC \Big(\big(\BX_1^d\big)\,\big({\BX_1^d})^{\rm T} \phi (\BX_1^d)\Big)
=\int\limits_{\bbR^d}
\bx\bx^{\rm T}\phi (\bx )f^{\rm{No}}_{\BC}(\bx )\rd\bx$$
and suppose that \eqref{3.3} is fulfilled. Then
\beq \hwphi(f)\leq \hwphi (f^{\rm{No}}_\BC),
\eeq
with equality iff $f=f^{\rm{No}}_\BC$ modulo $\phi$.
\et

The proof of Theorem \ref{Lemma 2.1} follows the argument in Example 3.1 from \cite{SY} repeated
verbatim in the multi-dimensional setting.

A conditional form of Theorem \ref{Lemma 2.1} is Theorem \ref{Lemma 2.2} below. The corresponding
assertion for the standard  entropy was noted in an earlier literature. See, e.g., Ref. [6, P. 1516]:
the proof of Theorem 29, item (c), the reference to a conditional version of [6, Lemma 5].
The proof of Theorem \ref{Lemma 2.2} is essentially hinted in its statement (see Eqn \eqref{eq:GmaxCond}),
and we omit it from the paper.

Given a $d\times d$ positive-definite matrix $\BC$ and  $p=1,\ldots ,d-1$, write $\BC$ in the
block form:
\beq\label{eq:BB}\BC =\begin{pmatrix}\BC_1^p&\BC_{n-p}^p\\ \BC_p^{n-p}&\BC_{p+1}^d\end{pmatrix}\eeq
where $\BC_{n-p}^p$ and $\BC_p^{n-p}$ are mutually transposed $p\times (n-p)$ and $(n-p)\times p$
matrices. Given $\bx=\left(\beac\bx_1^p\\ \bx_{p+1}^d\ena\right)$,
set $\BD\bx_{p+1}^d=\BC_{p}^{n-p}\;(\BC_{p+1}^d)^{-1} \bx_{p+1}^d$ and $\BK_1^p=\BC_1^p
-\BC_p^{n-p}\;(\BC_{p+1}^d)^{-1}\;\BC_{n-p}^p$. Correspondingly, if $\BX=\BX_1^d$ is a random vector
 (RV) with PDF $f_{\BX}$ and covariance matrix $\BC$ then
$\BC_1^p$ represents the covariance matrix for vector $\BX_1^p$, with PDF $f_{\BX_1^p}(\bx_1^p)$.
Let $\BX_{p+1}^d$
stand for the residual/remaining random vector and set $f_{\BX_{p+1}^d|\BX_1^p}(\bx_{p+1}^d|\bx_1^p)
=\diy\frac{f_\BX (\bx_1^d)}{f_{\BX_1^p}(\bx_1^p)}$. Also denote by $\BN$, $\BN_1^p$ and $\BN_{p+1}^d$
the corresponding Gaussian vectors, with PDFs $f_{\BN}(\bx )=f^{\rm{No}}_{\BC}(\bx )$, $f_{\BN_1^p}
(\bx_1^p)=f^{\rm{No}}_{\BC_1^p}(\bx_1^p)$ and
$f^{\rm{No}}_{\BN_{p+1}^d|\BN_1^p}(\bx_{p+1}^d|\bx_1^p)$. Finally, for a given WF $\bx\in\bbR^d
\mapsto \phi (\bx)$ set:
\beq\begin{array}{c}
\diy\psi (\bx_1^p)=\int\limits_{\bbR^{n-p}}\phi (\bx)f_{\BN_{p+1}^d|\BN_1^p}
(\bx_{p+1}^d|\bx_1^p)\rd\bx_{p+1}^d,\\
\diy\alpha (\BC_1^p) =\int\limits_{\bbR^p}\psi (\bx_1^p)f_{\BN_1^p}
(\bx_1^p)\rd\bx_1^p,\;\alpha (\BC) =\int\limits_{\bbR^d}\phi (\bx )f_\BN
(\bx )\rd\bx,\\
\diy\bPsi_{\BN_1^p}=\int\limits_{\bbR^p}\left[\bx_1^p\left(\bx_1^p\right)^{\rm T}\right]
\psi (\bx_1^p)f_{\BN_1^p}(\bx_1^p)\rd\bx_1^p,\;
\diy\bPhi_{\BN}=\int\limits_{\bbR^d}\left(\bx \bx^{\rm T}\right)
\phi (\bx)f_{\BN_1^d}(\bx )\rd\bx .
\end{array}\eeq
Also, consider inequalities
\beq\label{eq:Cnd1Gbbs}\beac\diy\int\limits_{\bbR^d}\phi (\bx )f_{\BX_1^p}(\bx_1^p)\Big[f_{\BX_{p+1}^d|
\BX_1^p}(\bx_{p+1}^d|\bx_1^p)
-f_{\BN_{p+1}^d|\BN_1^p}(\bx_{p+1}^d|\bx_1^p)\Big]\rd\bx\geq 0,\\
\beal\diy
\int\limits_{\bbR^d}\phi (\bx)\Big[f_{\BX}(\bx)-f_{\BN}(\bx)\Big]\bigg\{\diy\log\,\left[(2\pi )^p
{\rm{det}}\,\left(\BK_1^p\right)^{-1}\right]\\
\qquad\qquad\qquad\diy +(\log\,e)\left[
\left(\bx_1^p-\BD\bx_{p+1}^d\right)^{\rm T}\left(\BK_1^p\right)^{-1}\left(\bx_1^p-\BD\bx_{p+1}^d\right)\right]\bigg\}\rd\bx\leq 0.\ena\ena\eeq
\vskip .5 truecm

\bt\label{Lemma 2.2}
Make an assumption that bounds \eqref{eq:Cnd1Gbbs} are satisfied.
Then the following inequality holds true:
\beq\label{eq:GmaxCond}\begin{array}{l}\diy
\hwphi (\BX_{p+1}^d|\BX_1^p):=-\int\limits_{\bbR^d}\phi (\bx )f_\BX (\bx )\log\,f_{\BX_{p+1}^d|\BX_1^p}(\bx_{p+1}^d|\bx_1^p)\rd\bx\\
\qquad\qquad\qquad\;\;\leq\hwphi (\BN_{p+1}^d|\BN_1^p)=\hwphi (\BN)-\hw_{\psi}(\BN_1^p)\\
\diy\qquad =\frac{\alpha (\BC )}{2} \log \left[ (2\pi)^d {\rm{det}}\,\BC\right]+\frac{\log\,e}{2}{\rm{tr}}\,\left[\BC^{-1}\bPhi_\BN\right]\\
\qquad\qquad\qquad\qquad\diy-\frac{\alpha (\BC_1^p)}{2} \log \left[ (2\pi)^p {\rm{det}}\,\BC_1^p\right]-\frac{\log\,e}{2}{\rm{tr}}\,\left[\big(\BC_1^p\big)^{-1}\bPsi_{\BN_1^p}\right]\,.\end{array}\eeq
\et

\section{Weighted entropies under mappings}\label{sect:WEM} 

In this section we give  a series general theorems (Theorems \ref{lem:1.1} --  \ref{lem:1.3} and Theorem
\ref{lemma2:1})
 reflecting properties of the WEs under mappings of random variables (an example is a sum). Of a
 special importance for us is Theorem \ref{lem:1.3} used in Section 5. In essence, Theorems \ref{lem:1.1} --  \ref{lem:1.3} are repetitions of their counterparts from \cite{SY}, and
we omit their proofs.

\bt\label{lem:1.1} {\rm{(Cf. Lemma 1.1 from \cite{SY}.)}} Let $(\cX,\fX,\nu_\cX)$,
$(\cY,\fY,\nu_Y)$ be a pair of  Lebesgue spaces and suppose $X$, $Y$ are random elements
in $(\cX,\fX)$,  $(\cY,\fY)$  and PM/DFs
$f_X$, $f_Y$, relative to measures $\nu_\cX$, $\nu_\cY$, respectively. Suppose $\eta :(\cX,\fX)\to (\cY,\fY)$
is a measurable map onto, and that $\nu_\cY (B) =\nu_\cX(\eta^{-1}B)$,   $B\in\fY$.  Consider the partition
of $\cX$ with elements $\cB(y):=\{x\in\cX:\;\eta x=y\}$ and let $\nu_\cX (\,\cdot\,|y)$ be the family of
induced measures on $\cB(y)$, $y\in\cY$.
Suppose that $f_Y (y)=\diy\int\limits_{\cB (y)}f_X(x)\nu (\rd x|y)$ and for $x\in\cB (y)$ let $f_{X|Y}(x|y):=
\diy\frac{f_X(x)}{f_Y(y)}$ denote the PM/DF of $X$ conditional on $Y=y$. (Recall,
$f_{X|Y}(\,\cdot\,|y)$ is a family of PM/DFs defined for $f_Y$-a.a $y\in\cY$ such that
$\diy\int\limits_\cX G(x)f_X(x)\nu_\cX(\rd x)=
\int\limits_\cY\int\limits_{\cB (y)}G(x)f_{X|Y}(x|y)\nu_\cX(\rd x|y)f_Y(y)\nu_\cY(\rd y)$ for any non-negative
measurable function $G$.) Suppose that a WF\; $x\in\cX\mapsto\phi (x)\geq 0$ obeys
\beq\label{eq:1.10}\int\limits_\cX\phi (x)f_X(x)\Big[f_{X|Y}(x|\eta x)-1\Big]
\nu_\cX (\rd x)\leq 0\eeq
and set
\beq\psi (y)=\int\limits_{\cB(y)}\phi (x)f_{X|Y}(x|y)\nu (\rd x|y),\;\;y\in\cY.\eeq
Then
\beq\label{eq:1.11}\beal\diy\hwphi (X)\geq \hw_{\psi}(Y):=-\int\limits_\cY\psi (y)f_Y(y)\log\,f_Y(y)
\nu_Y(\rd y),\;\hbox{ or}\\ \diy\hwphi (X|Y):=-
\int\limits_\cX \phi (x)f_X(x)\log\,f_{X|Y}(x|y (x))\nu_\cX(\rd x)\geq 0,\ena\eeq
with equality iff \;$\phi (x)\big[f_{X|Y}(x|\eta x)-1\big]=0$\; for $f$-a.a. $x\in\cX$.

In particular, suppose that for $f_Y$-a.a. $y\in\cY$ set\; $\cB (y)$ contains at most countably many values and $\nu (\,\cdot\,|y)$ is a counting measure
with $\nu_1(x)=1$, $x\in\cB(y)$. Then the value $f_{X|Y}(x|\eta x)$ yields the conditional probability
$\bbP (X=x|Y=\eta x)$, which is $\leq 1$\; for \;$f_Y$-a.a. $y\in\cY$. Then $\hwphi (X|Y)\geq 0$ and the bound is strict unless, modulo $\phi$, map $\eta$ is $1-1$.
\et

\bt\label{lem:1.2} {\rm{(Cf. Lemma 1.2 from \cite{SY}.)}} Let $(\cX,\fX,\nu_\cX)$,
$(\cY,\fY,\nu_\cY)$, $(\cZ,\fZ,\nu_\cZ)$ be a triple of SMSs and suppose $X$, $Y$, $Z$ are random
elements in $(\cX,\fX)$,  $(\cY,\fY)$, $(\cZ,\fZ)$. Let $f_X$ be the PM/DF for $X$ relative to measure
$\nu_\cX$ and $f_{Y,Z}$ the joint PM/DF for $Y,Z$ relative to measures $\nu_\cY\times\nu_\cZ$.
Further, set $f_Z(z):=\diy\int\limits_\cY f(y,z)\nu_\cY(\rd y)$ and $f_{Y|Z}(y|z)=\diy\frac{f_{Y,Z}(y,z)}{f_Z(z)}$.
Suppose that
$$\eta :(\cX,\fX)\to (\cY,\fY),\;\;\zeta :(\cX,\fX)\to (\cZ,\fZ)$$
is a pair of measurable maps onto, and that
$$\nu_\cY (A) =\nu_\cX(\eta^{-1}A),\;A\in\fY,\;\;\nu_\cZ (B) =\nu_\cX(\zeta^{-1}B),\;B\in\fZ.$$
Consider the partition of $\cX$
with elements $\cB(y,z):=\{x\in\cX:\;\eta x=y,\zeta x=z\}$ and let $\nu_\cX (\,\cdot\,|y,z)$ be the family of induced measures on $\cB(y,z)$, $(y,z)\in\cY\times\cZ$.
Suppose that
$$f_{Y,Z} (y,z)=\diy\int\limits_{\cB (y,z)}f_X(x)\nu_\cX(\rd x|y,z)$$
and for $x\in\cB (y,z)$ let $f_{X|Y,Z}(x|y,z):=
\diy\frac{f_X(x)}{f_{Y,Z}(y,z)}$ denote the PM/DF of $X$ conditional on $Y=y$, $Z=z$. (Recall,
$f_{X|Y,Z}(\,\cdot\,|y,z)$ is a family of PM/DFs defined for $f_{Y,Z}$-a.a $(y,z)\in\cY\times\cZ$ such that
$$\diy\int\limits_\cX G(x)f_X(x)\nu_\cX(\rd x)=
\int\limits_{\cY\times\cZ}\int\limits_{\cB (y,z)}G(x)f_{X|Y,Z}(x|y,z)\nu_\cX(\rd x|y,z)f_{Y,Z}(y,z))\nu_\cY(\rd y)
\nu_\cZ(\rd z)$$
for any non-negative
measurable function $G$.) Assume that a WF\;$x\mapsto\phi (x)\geq 0$ obeys
\beq\label{eq:1.16}\int_{\cX}\phi (x)f(x)\Big[f_{X|Y,Z}(x|\eta x,\zeta x)-1\Big]\nu_\cX(\rd x)
\leq 0\eeq
and set
\beq\psi (y,z)=\int\limits_{\cB(y,z)}\phi (x)f_{X|Y,Z}(x|y,z)\nu (\rd x|y,z).
\eeq
Then
\beq \label{eq:1.17}\beal\diy -\int\limits_{\cY\times\cZ}\psi (y,z)f_{Y,Z}(y,z)\log\,f_{Y|Z}(y|z)\nu_\cY(\rd y)
\nu_\cZ(\rd z)\\
\qquad\qquad\diy=:\hw_{\psi}(Y|Z)\leq \hwphi(X|Z):=-\int\limits_{\cX}\phi (x)f_X(x)\log\,f_{X|Z}(x|\zeta x)
\nu (\rd x);\ena\eeq
equality iff \;$\phi (x)\big[f_{X|Y,Z}(x|\eta x,\zeta x)-1\big]=0$\; for $f_X$-a.a. $x\in\cX$.

As in Theorem {\rm\ref{lem:1.1}}, assume $\cB (y,z)$ consists of at most countably many values and
$\nu(x|y,z)=1$,
$x\in\cB (y,z)$ for $f_{Y,Z}$-a.a. $(y,z)\in\cY\times\cZ$. Then the value $f_{X|Y,Z}(x|y,z)$ yields the
conditional probability
$\bbP (X=x|Y=y,Z=z)$, for \;$f_{Y,Z}$-a.a. $y,z\in\cY\times\cZ$. Then $\hwphi (X|Z)\geq
\hw_\psi(Y|Z)$, with equality iff, modulo $\phi$, the map $x\mapsto (\eta x,\zeta x)$ is $1-1$.
\et

\bt\label{lem:1.3} {\rm{(Cf. Lemma 1.3 from \cite{SY}.)}} Let $(\cX,\fX,\nu_\cX)$,
$(\cY,\fY,\nu_\cY)$, $(\cZ,\fZ,\nu_\cZ)$ be a triple of SMSs and suppose $X$,
$Y$, $Z$ are random elements in $(\cX,\fX)$,  $(\cY,\fY)$, $(\cZ,\fZ)$. Let $f_{X,Y}$ be the joint PM/DF for
$X,Y$ relative to measure $\nu_\cX\times\nu_\cY$ and set
$$f_Y(y)=\int\limits_{\cX}f_{X,Y}(x,y)\nu_\cX (\rd x),\;\;f_{X|Y}(x|y)=\frac{f_{X,Y}(x,y)}{f_Y(y)}.$$
Suppose that
$$\xi :(\cY,\fY)\to (\cZ,\fZ)$$
is a measurable maps onto, and that
$$\nu_\cZ (C) =\nu_\cY(\xi^{-1}C),\;C\in\fZ.$$
Consider a partition of $\cY$  with elements $\cC(z):=\{y\in\cY:\;\xi y=z\}$ and let $\nu_\cY (\,\cdot\,|z)$
be the family of induced measures on $\cC (z)$, $z\in\cZ$.
Given $(x,z)\in\cX\times\cZ$ and $y\in\cC (z)$, let
$$f_{X,Z} (x,z)=\diy\int\limits_{\cC (z)}f_{X,Y}(x,y)\nu_\cY(\rd y|z),\;f_Z(z)=\int\limits_{\cX}f_{X,Z}(x,z)
\nu_\cX (\rd x),$$
and
$$f_{X|Z}(x|z)=\frac{f_{X,Z}(x,z)}{f_Z(z)},\;\;f_{Y|Z}(y|z)=\frac{f_Y(y)}{f_Z(z)}.$$
Assume that a WF\;$(x,y)\mapsto\phi (x,y)\geq 0$ obeys
\beq\label{eq:1.161}\diy\int_{\cX\times\cY}\phi (x,y)\left[f_{X,Y}(x,y)-f_Z(\xi y)f_{X|Z}(x|\xi y)f_{Y|Z}(y|\xi y)
\right]\nu_\cX(\rd x )\nu_\cY(\rd y )\geq 0\eeq
and set
\beq\psi (x,z)=\int\limits_{\cC(z)}\phi (x,y)f_{Y|Z}(y|z)\nu_\cY(\rd y|z).
\eeq
Then
\beq \label{eq:1.171}\beal\diy -\int\limits_{\cX\times\cZ}\psi (x,z)f_{X,Z}(x,z)\log\,f_{X|Z}(y|z)\nu_\cX(\rd x)
\nu_\cZ(\rd z)\\
\qquad\diy=:\hw_{\psi}(X|Z)\geq \hwphi(X|Y):=-\int\limits_{\cX\times\cY}\phi (x,y)f_X(x)\log\,f_{X|Y}(x|y)
\nu_\cX (\rd x)\nu_\cY (\rd y).\ena\eeq
Furthermore, equality in \eqref{eq:1.171} holds iff  $X$ and $Y$ are conditionally independent given
$Z$ modulo $\phi$, i.e. $\phi (x,y)\left[f_{X,Y}(x,y)-f_Z(\xi y)f_{X|Z}(x|\xi y)f_{Y|Z}(y|\xi y)\right]=0$.
\et

We will use an alternative notation $\hwphi (\BX):=\hwphi (f_\BX)$ where $\BX=\BX_1^d =
\begin{pmatrix}X_1\\ \vdots\\ X_d\end{pmatrix}$ is
a $d$-dimensional random vector with PDF $f_\BX (\bx)$. 
In this context, we employ the notation $\BX\sim f_\BX$, $\BY\sim f_\BY$, $(\BX,\BY)\sim f_{\BX,\BY}$
and $(\BX|\BY)\sim f_{\BX|\BY}$ where $f_{\BX|\BY}(\bx|\by)=\diy\frac{f_{\BX,\BY}(\bx,\by )}{f_\BY(\by )}$.

Theorem \ref{lemma2:1} below mimics a result in \cite{CT}, extending from the case
of a standard entropy to that of the WE. A number of  facts are related to the conditional WE
$$\hwphi (\BX|\BY):=-\diy\int_{\bbR^d\times\bbR^d}\phi (\bx,\by)f_{\BX,\BY}(\bx,\by )\log\,f_{\BX|\BY}(\bx|\by)
\rd\bx\rd\by$$
or, more generally,
$$\hw_{\wt\phi} (\BU|\BV):=-\diy\int_{\bbR^d\times\bbR^d}{\wt\phi}(\bu,\bv)f_{\BU,\BV}(\bu,\bv )
\log\,f_{\BU|\BV}(\bu|\bv)\rd\bu\rd\bv,$$
Here a pair $(\BU,\BV)$  is a function of $(\BX,\BY)$ with a joint PM/DF $f_{\BU,\BV}$, marginal
PM/DFs $f_{\BU}$, $f_{\BV}$ and conditional PM/DF $f_{\BU|\BV}(\bu|\bv):=\diy \frac{f_{\BU,\BV}
(\bu,\bv)}{f_{\BV}(\bv)}$. (Viz., $\BU=\BY$, $\BV=\BX+\BY$.) WF ${\wt\phi}$ may or may not be
involved with the map $(\BX,\BY)\mapsto (\BU,\BV)$.

\bt\label{lemma2:1}
Suppose $\BX$ and $\BY$ are independent random vectors of
dimension $d$, with PDFs $f_\BX$ and $f_\BY$:
$$(\BX,\BY)\sim f_{\BX,\BY}\;\hbox{ where }\;f_{\BX,\BY}(\bx,\by )=f_\BX (\bx )f_\BY(\by ), \;\;\bx,\by\in\bbR^d.$$
Assume that WF $\;(\bx,\by)\in \bbR^d\times \bbR^{d^\prime}
\rightarrow \phi(\bx,\by)\geq 0$ obeys
\beq\label{eq:sumcond} \int\limits_{\bbR^d\times\bbR^d}\phi(\bx,\by)f_\BY (\by)\Big[ f_{\BX}(\bx)-
f_{\BX+\BY}(\bx+\by)\Big]\rd\bx\rd\by\geq 0\eeq
and set
\beq\label{eq:zetas}\theta (\bv )=\int\limits_{\bbR^d}\phi(\bv-\by,\by)f_{\BY|\BX+\BY}(\by|\bv )\rd\by,\;
\theta^*(\bx)=\int\limits_{\bbR^d}\phi (\bx+\by,\by)f_\BY(\by)\rd\by,\;\;\bv,\bx\in\bbR^d.\eeq
Then
\beq\label{eq:sumbnd}h^{\rm w}_\theta (\BX+\BY)\geq h^{\rm w}_{\theta^*}(\BX),\eeq
with equality iff $\phi (\bx,\by)f_\BY (\by)\Big[ f_{\BX}(\bx)-
f_{\BX+\BY}(\bx+\by)\Big]=0$ for Lebesgue-a.a. $(\bx,\by)\in\bbR^d\times\bbR^d$.
\et

\bp
Set: $\phi^*(\bx ,\by )=\phi (\bx+\by ,\by)$. The following relations (a)--(c) hold true:
\beq\beac{\rm{(a)}}\quad h^{\rm w}_{\theta}(\BX+\BY)\geq \hwphi(\BX+\BY|\BY),\\
{\rm{(b)}}\quad\hwphi (\BX+\BY|\BY)=h^{\rm w}_{\phi^*}(\BX|\BY),\;\;{\rm{(c)}}\quad h^{\rm w}_{\phi^*}
(\BX|\BY)= h^{\rm w}_{\theta^*}(\BX).\ena\eeq
Here bound (a) comes from the sub-additivity of the WE, see \cite{SY}, Theorem 1.3 or Eqn (1.31) from \cite{SY}. Next, (b) is
derived by applying the following equations:
\beqq \hwphi(\BX+\BY|\BY)&=&\int\limits_{\bbR^d}f_\BY(\by) \hwphi (\BX+\BY|\BY=\by)\rd\by\\
&=&-\int\limits_{\bbR^d\times\bbR^d}\phi (\bx+\by,\by) f_\BY(\by) f_{\BX|\BY}(\bx|\by)\log f_{\BX|\BY}
(\bx|\by)\rd\bx\rd\by.\qquad\eeqq
Finally, Eqn (c) holds because $\BX$ and $\BY$ are independent.

The proof of Theorem \ref{lemma2:1} is completed by observing that
\beqq h^{\rm w}_{\phi^*}(\BX|\BY)&=&-\int\limits_{\bbR^d\times\bbR^d}\phi(\bx+\by,\by)f_{\BX,\BY}
(\bx,\by)\log f_{\BX|\BY}(\bx|\by)\rd\bx)\rd\by\\
&=&-\int\limits_{\bbR^d}\left[\int\limits_{\bbR^d}\phi(\bx+\by,\by)f_\BY(\by)\rd\by\right]f_\BX(\bx)\log f_\BX(\bx)\rd\bx .
\qquad\qquad \eeqq
\ep

\br {\rm The assertion of Theorem {\rm{\ref{lemma2:1}}} remains valid, {\it mutates mutandis}, when
$\BX$ and $\BY$ have different dimensions. Viz., we can assume that $\BY$ has dimension $d^\prime <d$
and append $\BY$ and $\by$ with zero entries when we sum $\BX+\BY$ and $\bx+\by$.}  
\er

\section{Miscellaneous weighted determinant inequalities}\label{sect:MWDIs}

In this section we present  a host of WDIs derived from properties of the WEs.
As we said before, the proposed inequalities hold when WF $\phi \equiv 1$
(in this case the stated conditions are trivially fulfilled). To stress parallels with
`standard' DIs, we provide references to \cite{CT} or \cite{DCT} in each case under
consideration.

\bt
{\rm{(Cf. \cite{CT} Theorem 2.)}} Let $\BX$, $\BY$ be independent $d$-variate normal vectors with zero means and covariance matrices $\BC_1$, $\BC_2$, respectively: $f_{\BX,\BY}(\bx,\by )=f_\BX(\bx )f_{\BY}(\by )$, $\bx,\by\in\bbR^d$, where
$f_\BX=f^{\rm{No}}_{\BC_1}$, $f_\BY=f^{\rm{No}}_{\BC_2}$. Given a WF  $(\bx,\by )\in \bbR^d\times\bbR^d\mapsto \phi(\bx,\by)\geq 0$, positive on an open domain in $\bbR^d\times\bbR^d$, consider a quantity $\beta$ and $d\times d$ matrices $\bTha$, $\bTha^*$:
\beq\beta=\int\limits_{\bbR^d}\theta (\bx)f^{\rm{No}}_{\BC_1+\BC_2}(\bx)\rd\bx,\;  
\bTha =\int\limits_{\bbR^d}\bx\bx^{\rm T}\theta (\bx)f^{\rm{No}}_{\BC_1+\BC_2}(\bx)\rd\bx ,\;
\bTha^*=\int\limits_{\bbR^d}\bx\bx^{\rm T}\theta^*(\bx)f^{\rm{No}}_{\BC_1}(\bx)\rd\bx\eeq
where $\theta$ and $\theta^*$ are as in \eqref{eq:zetas}:
\beq\label{eq:zetass}\theta (\bx)=\int\limits_{\bbR^d}\phi(\bz,\bx-\bz)f_{\BY|\BX+\BY}(\bx-\bz|\bx)\rd\bz,\;\;
\theta^*(\bx)=\int\limits_{\bbR^d}\phi (\bx+\by,\by)f_\BY(\by)\rd\by.\eeq
Assume the  condition emulating \eqref{eq:sumcond}:
\beq\label{eq:sumcondi} \diy\int\limits_{\bbR^d\times \bbR^d}\phi(\bx,\by)f^{\rm{No}}_{\BC_2}(\by)\Big[f^{\rm{No}}_{\BC_1}(\bx)-f^{\rm{No}}_{\BC_1+\BC_2}(\bx +\by)\Big]\rd\bx\rd\by \geq 0.\eeq
Then
\beq \label{eq2:1}\beta\log\left[\frac{{\rm det}\,(\BC_1+\BC_2)}{{\rm det}\,\BC_1}\right]+(\log e) \left\{{\rm tr}\Big[(\BC_1+\BC_2)^{-1}\bTha\Big]-{\rm{tr}}\left(\BC_1^{-1}\bTha^*\right)\right\}\geq 0.\eeq
\et

\bp Using Theorem \ref{lemma2:1} and Eqn \eqref{eq:sigm}, we can write:
\beqq \begin{array}{l} \diy \frac{1}{2} \log \left[ (2\pi)^d ({\rm{det}}\,(\BC_1+\BC_2))\right]
\int\limits_{\bbR^d}\theta (\bx )f^{\rm{No}}_{\BC_1+\BC_2}(\bx )\rd\bx +\frac{\log\,e}{2}{\rm{tr}}\,(\BC_1
+\BC_2)^{-1}\bTha\\
\qquad\qquad\qquad\qquad\diy \geq\frac{1}{2}\, \log \left[ (2\pi)^d ({\rm{det}}\,\BC_1)\right]
\int\limits_{\bbR}\theta^*(\bx )f_{\BC_1}(\bx )\rd\bx +\frac{\log\,e}{2}{\rm{tr}}\,\BC_1^{-1}\bTha^*.
\end{array}\eeqq
The bound in \eqref{eq2:1} then follows.
\ep

\br {\rm It is instructive to observe that \eqref{eq2:1} is equivalent to:
\beq \label{eq2:2}\beal\diy\beta\log\left[{\rm det}\,(\BI+\BC^{-1}_1\BC_2)\right]\\
\qquad\qquad +(\log e) {\rm tr}\left[(\BC_1+\BC_2)^{-1}\bTha^*-\BC_1^{-1}\bTha^*+(\BC_1+\BC_2)^{-1}\wt{\bTha}\right]\geq 0\ena\eeq
where
\beqq\wt{\bTha}=\diy\int\limits_{\bbR^d\times\bbR^d}\left(\bx\by^{\rm T}+\by\bx^{\rm T}+\by\by^{\rm T}\right)
\phi(\bx +\by ,\by )f^{\rm{No}}_{\BC_2}(\by )f^{\rm{No}}_{\BC_1}(\bx )\rd\by\rd\bx .\eeqq
This claim is verified by observing that $\bTha =\bTha^*+\wt{\bTha}$.}
\er

\br {\rm As above, we can assume that $\BC_2$ is a matrix of size $d^\prime\times d^\prime$,
agreeing that
in the sum $\BC_1+\BC_2$, matrix $\BC_2$ is identified as a top left block (say). This is possible because
in Eqns \eqref{eq2:1} and \eqref{eq2:2} we do not use the inverse $\BC_2^{-1}$ or the determinant ${\rm{det}}\,\BC_2$.}
\er

To this end, recall the following theorem from \cite{Mi}:

\bt  Let $\BG$ and $\BG+\BE$ be nonsingular matrices where $\BE$ is a matrix of rank one. Let $g={\rm{tr}}\,\left( \BE \BG^{-1}\right)$. Then $g\neq -1$ and
\beqq (\BG+\BE)^{-1}=\BG^{-1}-\frac{1}{1+g}\BG^{-1}\;\BE\;\BG^{-1}.\eeqq
\et

The above equation is essentially the Sherman-Morrison formula (see \cite{DB}, p. 161).\\

Assuming that $\BC_2=\BE$ has rank 1 and letting $g={\rm tr}\,(\BE\BC_1^{-1})$, inequality
(\ref{eq2:1}) turns into the following bound:
\beq \label{eq2:3}\beta\log\left[\frac{{\rm det}\,(\BC_1+\BE)}{{\rm det}\,\BC_1}\right]+(\log e)\left[-{\rm tr}\left(\frac{\BC_1^{-1}\BE\BC_1}{1+g}\bTha^*\right)+{\rm{tr}}\,\{(\BC_1+\BE)^{-1}\wt{\bTha}\}\big)\right]\geq 0.\eeq

The techniques developed so far allows us to prove Theorem \ref{thm:extSzasz} below rendering
a weighted form
of Szasz theorem. Suppose $\BC$ is a positive definite $d\times d$ matrix. Given $1\leq k\leq d$
and a set $S\subseteq I^{(d)}:=\{1,\ldots ,d\}$ with $\#(S)=k$, denote by $\BC(S)$ be the $k\times k$ sub-matrix of $\BC$ formed by the rows and columns with indices $i\in S$.
With every $S$ we associate a Gaussian random vector $\BX (S)\sim f^{\rm{No}}_{\BC (S)}$ considered
as a sub-collection of $\BX\sim f^{\rm{No}}_{\BC}$. Accordingly, conditional PDFs emerge,
$f^{\rm{No}}_{S|S^\prime}(\bx(S)|\bx (S^\prime ))$, for pairs of sets $S,S^\prime$ with $S\cap S^\prime=\emptyset$, where $\bx (S)\in\bbR^{\# (S)}$, $\bx (S^\prime )\in\bbR^{\# (S^\prime)}$. [The PDF  $f^{\rm{No}}_{S|S^\prime}$
is expressed in terms of block sub-matrices forming the inverse matrix $\BC(S\cup S^\prime )^{-1}$.]

Further, let a function $\phi (\bx )\geq 0$, $\bx\in\bbR^d$, be given, which is positive on an open domain in
 $\bbR^ d$ and set, as in \eqref{eq:psiS},
 \beq\label{eq:psiSN} \psi (S;\bx (S))=\int\limits_{\bbR^{\#(\oS)}} \phi (\bx)f^{\rm{No}}_{\oS|S}(\bx(\oS)\,|\,\bx(S))\rd\bx(\oS).\eeq
Furthermore, define:
\beq\label{eq:tauT} \tau (S)={\rm tr}\;\big[\BC(S)^{-1}\bPhi (S)\big],\;\;\rT (k)=\sum\limits_{S\subseteq I^{(d)}:\# (S)=k}\tau (S)\eeq
where matrix $\bPhi (S)$ is given by
\beq\label{eq:bPhiSs} \bPhi (S)=\bPhi (\BC (S))=\int\limits_{\bbR^{\# (S)}} \bx (S)\bx (S)^{\rm T} \psi (S;\bx (S) ) f^{\rm{No}}_{\BC (S)}(\bx (S) )\rd\bx (S). \eeq
(For $S=I^{(d)}$, we write simply $\bPhi$; cf. \eqref{eq:bPhi}.) Finally, set:
\beq\label{eq:alphaA}\alpha (S)=\alpha (\BC (S))=\int\limits_{\bbR^{\# (S)}} \psi (S;\bx (S)) f^{\rm{No}}_{\BC (S)}(\bx (S))\rd \bx (S),\;\;\rA (k)=\sum\limits_{S\subseteq I^{(d)}:\# (S)=k}\alpha (S)\eeq
and
\beq\label{eq:lamL}\lam (S)=\alpha (S)\log\,{\rm{det}}\,\BC(S),\;\;\Lam (k):=\sum\limits_{S\subseteq I^{(d)}:\# (S)=k}\lam (S). \eeq

Consider the following condition invoking broken dependence and analogous to \eqref{eq:condSi}:
\beq\label{eq:cndGSiN}\beal\hbox{$\forall$ $i\in S\subseteq I$,
with $S^-_i=\{j\in S:\;j<i\}$ and $S^+_i=\{j\in S:\;j>i\}$,}\\
\diy\int\limits_{(\bbR^{\# (S)}}\psi (S;\bx (S))\Big\{f^{\rm{No}}_{\BC (S)}(\bx (S))\\
\quad -
f^{\rm{No}}_{\BC (S^-_i)}(\bx (S^-_i))\times\Big[f^{\rm{No}}_{\{i\}|S^-_i}(x_i|\bx (S^-_i))
f^{\rm{No}}_{S^+_i|S^-_i}(\bx (S^+_i)|\bx (S^-_i))\Big]\Big\}\rd\bx (S)\geq 0,\ena\eeq

\bt\label{thm:extSzasz}
{\rm{(Cf. \cite{CT}, Theorem 4 or \cite{DCT}, Theorem 31)}} Assume condition {\rm{\eqref{eq:cndGSiN}}}.
Then the quantity $m(k)=m(k,\BC,\phi )$ defined by
\beqq m(k):=\binom{d}{k}^{-1}\left[\frac{\log\;\Lam (k)}{k}+ \frac{\log\,(2\pi )}{2}
A(k)+\frac{\log\; e}{2k}\Lam (k)\right]\eeqq
is decreasing in $k=1,\ldots ,d$:
\beq\label{eq:m} m(1)\geq \ldots \geq m(d).\eeq
\et

\bp For $\BX(S)\sim f^{\rm{No}}_{\BC (S)}$ we have, by using  \eqref{eq:sigm}:
\beqq\frac{h^{\rm w}_{\psi(S)}(\BX(S))}{k}=\frac{\alpha (S)}{2k}\log\; \big[(2\pi)^k {\rm det}\,\BC (S)\big]
+\frac{\log e}{2k}{\rm tr}\left[\BC (S)^{-1}\bPhi{(S)}\right].\eeqq
Therefore,
\beqq  m(k) =\binom{d}{k}^{-1} \diy\sum\limits_{S:|S|=k}\left\{ \frac{\alpha (S)}{2k}\log\,\left[(2\pi)^k
{\rm det}\,\BC (S)\right]+\frac{\log\; e}{2k} {\rm tr}\,\left(\BC (S)^{-1}\bPhi (S)\right)\right\}\eeqq
Invoking Theorem \ref{average.WE} completes the proof.
\ep

\bt
{\rm{(Cf. \cite{CT}, Theorem 5 or \cite{DCT}, Theorem 32)}} Assuming  {\rm{\eqref{eq:cndGSiN}}}, for all $r>0$ the values
\beqq s(k)=s(k,\BC,\phi ):=\binom{d}{k}^{-1}\sum\limits_{S\subseteq I^{(d)}:\;\# (S)=k}\Lam (k)^{1/k}
\exp \bigg\{r\bigg[ \frac{\log\,(2\pi )}{2}
A(k)+\frac{\log\; e}{2k}\rT (k)\bigg] \bigg\}\eeqq
obey
\beq\label{eq:s} s(1)\geq \ldots\geq s(d).\eeq
\et

\bp The assertion follows readily from Theorem \ref{lemma.g}.
\ep

Our next goal is to establish bounds for Toeplitz determinants extending Theorem 6 from \cite{CT} (or Theorem 27 from \cite{DCT}).
It is said that $\BC =(C_{ij})$ is a $d\times d$ Toeplitz matrix if $C_{ij}=C_{kl}$ whenever $|i-j|=|k-l|$.
A more restrictive property is cyclic Toeplitz  where $C_{ij}=C_{kl}$ whenever ${\rm{dist}}_d(i,j)
={\rm{dist}}_d(k,l)$. Here, for $1\leq i<j\leq d$ the cyclic distance ${\rm{dist}}_d(i,j)=\min\,[j-i,d-j+i]$;
it is then extended to a metric with ${\rm{dist}}_d(i,j)={\rm{dist}}_d(j,i)$ and ${\rm{dist}}_d(i,i)=0$.
As before, we consider sub-matrices $\BC (S)$ where $S\subseteq I^{(d)}:=\{1,\ldots ,d\}$ and the Gaussian
random vectors $\BX (S)\sim f^{\rm{No}}_{\BC (S)}$ as sub-collections in $\BX_1^d:=\begin{pmatrix}X_1\\ \vdots\\ X_d\end{pmatrix}\sim f^{\rm{No}}_{\BC}$.
A special role is played by $S=I_{i,j}$ where $I_{i,j}$ stands for
a segment of positive integers $\{i,i+1,\ldots ,j\}$ of cardinality $j-i+1$ where $1\leq i<j\leq d$. In particular,
for $S=I_{1,k}$, we set: $\BC (S)=\BC_k$ and deal with vectors $\BX_1^k\sim f^{\rm{No}}_{\BC_k}$, $1\leq k\leq d$, with $\BC_d=\BC$.

Accordingly, we say that WF $\bx\in\bbR^d\mapsto \phi (\bx)\geq 0$ has a Toeplitz property if the value of the reduced WF $\psi (I_{i,j};\bx_i^j)$
coincides with $\psi (I_{i+k,j+k};\bx_{i+k}^{j+k})$, provided that arguments $\bx_i^j=\bx (I_{i,j})$ and $
\bx_{i+k}^{j+k}=\bx (I_{i+k,j+k})$ are
shifts of each other, where $1\leq i<j\leq d$ and $1\leq i+k<j+k\leq d$.  An example is where $\BC$ is cyclic Toeplitz and $\phi$ has
a product-form:  $\phi (\bx )=\prod\limits_{1\leq i\leq d}\varphi (x_i)$.  Recall, the reduced WF  in question involves the conditional PDF $f^{\rm{No}}_{I^{\ct}_{i,j}|I_{i,j}}(\bx (I^{\ct}_{i,j})|\bx_i^j)$:
$$\psi (I_{i,j};\bx_i^j)=\int\limits_{\bbR^{d-j+i-1}}\phi (\bx )f^{\rm{No}}_{I^{\ct}_{i,j}|I_{i,j}}(\bx (I^{\ct}_{i,j})|\bx_i^j)\rd\bx (I^{\ct}_{i,j})\;
\hbox{ where }\;I^{\ct}_{i,j}=I_{1,d}\setminus I_{i,j}.$$
For $S=I_{1,k}$, $1\leq k\leq d$, in accordance with  \eqref{eq:sigm},
\beq\label{eq:hwCk} h^{\rm}_{\psi (k)}(\BX_1^k)= h^{\rm}_{\psi (I_{1,k})}(\BX_1^k)=\diy \frac{\alpha (\BC_k)}{2}\;\log\;\left[(2\pi)^k  {\rm{det}}\,\BC_k\right]+\frac{\log e}{2}{\rm tr}\left[\BC_k^{-1}\bPsi_k\right].\eeq
Here the value $\alpha (\BC_k)=\alpha (\BC_k,\BC ,\phi )$ and the $k\times k$ matrix $\bPsi_k=\bPsi_k (\BC_k,\BC,\psi )$ are given by
\beq\label{eq:alphaCk}\alpha (\BC_k)=\int\limits_{\bbR^k} \psi (k;\bx_1^k) f^{\rm{No}}_{\BC_k}(\bx_1^k)\rd \bx_1^k,\;\;\bPsi_k =\int\limits_{\bbR^k} \bx_1^k\left(\bx_1^k\right)^{\rm T} \psi (k;\bx_1^k) f^{\rm{No}}_{\BC_k}(\bx_1^k)\rd\bx_1^k\eeq
and $\psi (k)=\psi (I_{1,k})$.  (For $k=d$, the subscript $k$ will be omitted.)



\bt
{\rm{(Cf. \cite{CT}, Theorem 6 or \cite{DCT}, Theorem 27)}} Suppose $\BC_n$ is a positive definite $d\times d$ Toeplitz
matrix and $\phi$ has the Toeplitz property. Consider the map
$k\in\{1,\ldots ,d\}\mapsto a(k)=a(k,\BC,\phi )$ where
\beq a(k)=\alpha (C_k)\left\{\log (2\pi )+\log\; \left[({{\rm{det}}\;\BC_k})^{1/k}\right]\right\}
+\frac{\log e}{k}{\rm tr}\left[\BC_k^{-1}\bPsi_k\right].\eeq
Assuming condition \eqref{eq:cndGSiN}, the value $a(k)$ is decreasing in $k$: $a(1)\geq\ldots\geq a(d)$.
\et

\bp 
By using the Toeplitz property of $\BC$ and $\phi$, we can write
\beq \label{equal.Toeplitz} h^{\rm w}_{\psi (I_{1,k})}(X_k|\BX_1^{k-1})=h^{\rm w}_{\psi (I_{2,k+1})}(X_{k+1}|\BX_2^k).\eeq
Next, Theorem \ref{lem:1.3} 
yields:
\beq \label{subadd.cwe} h^{\rm w}_{\psi (I_{2,k+1})}(X_{k+1}|\BX_2^k) \geq h^{\rm w}_{\psi (I_{1,k+1})}(X_{k+1}|\BX_1^k).\eeq

From (\ref{equal.Toeplitz}) and (\ref{subadd.cwe}) we conclude that $h^{\rm w}_{\psi (I_{1,k})}(X_k|\BX_1^{k-1})$ is decreasing in $k$. Thus the running average also decreases. On the other hand, by the chain rule
\beqq \frac{1}{k}\;h^{\rm}_{\psi (I_{1,k})}(\BX_1^k)=\diy\frac{1}{k}\sum\limits_{i=1}^k h^{\rm w}_{\psi (I_{1,i})}(X_i|\BX_1^{i-1}).\eeqq
Consequently $\diy\frac{1}{k}\; h^{\rm}_{\psi (I_{1,k})}(\BX_1^k)$ too decreases in $k$.
Referring to Eqns \eqref{eq:alphaCk} and \eqref{eq:hwCk}  leads directly to the result.
\ep

\bt
{\rm{(Cf. \cite{DCT}, Theorem 33.)}} Given a WF $\bx=\begin{pmatrix}x_1\\ \vdots\\ x_d\end{pmatrix}\in\bbR^d\mapsto\phi (\bx)$, assume condition
\beq\label{eq:Hada} \int\limits_{\bbR^d}\phi(\bx )\left[f^{{\rm{No}}}_C(\bx )-\prod\limits_{i=1}^n f^{{\rm{No}}}_{C_{ii}}(x_i)\right] \rd \bx \geq 0.\eeq
Then the quantity
\beqq \begin{array}{l} w(k)=w(k,\BC,\phi)=\diy \binom{d}{k}^{-1}\frac{\alpha(\BC)}{2k}\log\;
\left[\prod\limits_{S\subseteq I_n:\;\# (S)=k}\frac{(2\pi)^d ({{\rm{det}}\;\BC})}{(2\pi)^{d-k}({{\rm{det}}\;\BC(S^{\ct})})}\right]\\
\quad\quad +\diy\binom{d}{k}^{-1}\frac{\log\;e}{2k}\sum\limits_{S\subseteq I_n:\;\# (S)=k}\left\{{\rm tr}\left[\BC^{-1}\bPhi\right]-{\rm tr}\left[\BC(S^{\ct})^{-1}\bPhi(S^{\ct})\right]\right\}\end{array}\eeqq
is increasing in $k$, with
\beq\label{eq:w}w(1)\leq\dots\leq w(d).\eeq
\et

\bp Using the conditional WE, we can write
\beqq \hwphi(\BX(S)|\BX(S^{\ct}))&=& \hwphi(\BX(S),\BX(S^{\ct}))-h^{\rm w}_{\psi(S^{\ct})}(\BX(S^{\ct}))\\
&=&\diy\frac{\alpha(\BC)}{2} \log\;\left[(2\pi)^d({\rm{det}}\BC)\right]+\frac{\log\;e}{2}{\rm tr}\left[\BC^{-1}\bPhi\right]\\
&-&\diy\frac{\alpha(\BC)}{2} \log\;\left[(2\pi)^{d-k}({\rm{det}}\BC(S^{\ct}))\right]
+\frac{\log\;e}{2}{\rm tr}\left[\BC(S^{\ct})^{-1}\bPhi(S^{\ct})\right].\eeqq
Here $\alpha(\BC)=\int\limits_{\bbR^d}\phi(\bx )f^{{\rm{No}}}_{\BC}(\bx )\rd \bx
=\int\limits_{\bbR^{\#(S^{\ct})}}\psi(\bx(S^{\ct}))f^{{\rm{No}}}_{{\BC}(S^{\ct})}(\bx(S^{\ct}))\rd \bx(S^{\ct})$.
Therefore,
\beq\label{CWEGaussian}\beal \hwphi(\BX(S)|\BX(S^{\ct}))\\
\quad\diy=\frac{\alpha(\BC)}{2}\log\; \left[\frac{(2\pi)^d ({{\rm{det}}\;\BC})}{(2\pi)^{d-k}({{\rm{det}}\;
\BC(S^{\ct})})}\right]
+\frac{\log\;e}{2} \left\{{\rm tr}\left[\BC^{-1}\bPhi\right]-{\rm tr}\left[\BC(S^{\ct})^{-1}\bPhi(S^{\ct})
\right]\right\}.\ena\eeq
After that we apply Theorem \ref{average.CWE} which completes the proof.
\ep

\br {\rm Note that the outermost inequality, $w(1)\leq w(d)$, can be rewritten as
\beq\label{eq:w1wd} \beal \alpha(\BC)\log\;\left[(2\pi)^d ({\rm{det}}\,\BC)\right]+\log\;e\;{\rm tr}\left[\BC^{-1}\bPhi\right]\geq
\diy\alpha(\BC)\log \;\left[\prod\limits_{i=1}^d \frac{2\pi({\rm {det}}\;\BC)}{{\rm{det}}\;
\BC(I_1^{i-1}\cup I_{i+1}^d)}\right]\\
\qquad\qquad\qquad\qquad
\diy+\log\;e\sum\limits_{i=1}^d \left\{{\rm tr}\left[\BC^{-1}\bPhi\right]-{\rm tr}\left[\BC(I_1^{i-1}\cup
I_{i+1}^d)^{-1}\bPhi(I_1^{i-1}\cup I_{i+1}^d)\right]\right\}.\ena\eeq}
\er
\vskip .5 truecm

Our next goal is to establish additional WDIs  by using Theorem \ref{averagemutualWE2}. For
this purpose, we first analyse the mutual Gaussian WE,
$i^{\rm w}_{\phi}(\BX(S):\BX(S^{\ct}))$.
According to the definition of the mutual WE in \cite{SY}, we can write
\beqq i^{\rm w}_{\phi}(\BX(S):\BX(S^{\ct}))=h^{\rm w}_{\psi(S)}(\BX(S))-\hwphi(\BX(S)|\BX(S^{\ct})).
\eeqq
Then, in accordance with (\ref{CWEGaussian}), we have
\beqq \begin{array}{l}\diy i^{\rm w}_{\phi}(\BX(S):\BX(S^{\ct}))=\frac{\alpha(\BC)}{2} \log\;\left[\frac{({\rm
{det}}\BC(S))\;({\rm {det}}\BC(S^{\ct}))}{({\rm {det}}\BC)}\right]\\
\quad\quad +\diy \frac{\log\;e}{2}\left\{{\rm tr}\left[\BC(S)^{-1}\bPhi(S)\right]+{\rm tr}\left[
\BC(S^{\ct})^{-1}\bPhi(S^{\ct})\right]-{\rm tr}\left[\BC^{-1}\bPhi\right]\right\}.\end{array}\eeqq

In Theorems \ref{thm:u} and \ref{thm:z} we consider the following condition \eqref{cond.MWEN} stemming
from \eqref{cond.MWE}:
$\forall$ $S\subseteq\{1,\ldots ,n\}$ with $\#\,S\geq 2$ and $i,j\in S$ with $i\neq j$,
\beq\label{cond.MWEN}\beal\diy\int\limits_{\bbR^d}\phi(\bx) f^{\rm{No}}_{\oS|S}(\bx(\oS)|\bx (S))
\Big[f^{\rm{No}}_{\BC (S)}(\bx(S))\\
\quad -f^{\rm{No}}_{\BC (S\setminus\{i,j\})}(\bx (S\setminus\{i,j\})\;f^{\rm{No}}_{i|S\setminus\{i,j\}}(\bx (S
\setminus\{i,j\})\;f^{\rm{No}}_{j|S\setminus\{i,j\}}(x_j|
\bx (S\setminus\{i,j\})\Big]\rd\bx \geq 0.\ena\eeq
The proof of Theorems \ref{thm:u} and \ref{thm:z} is done with the help of Theorem
\ref{averagemutualWE2},
assuming that $X_1,X_2,\dots,X_d$ are normally distributed with covariance matrix $\BC$.




\bt\label{thm:u}
{\rm{(Cf. \cite{DCT}, Theorem 34.)}}\; Assume condition \eqref{cond.MWEN}. Let
\beqq \begin{array}{l} u(k)=\diy\binom{d}{k}^{-1}\frac{\alpha(\BC)}{2k} \log\;\left[
\prod\limits_{S\subseteq I^{(d)}:\;\# (S)=k}\frac{({\rm {det}}\BC(S))\;({\rm {det}}\BC(S^{\ct}))}{({\rm {det}}\BC)}
\right]\\
\qquad +\diy\binom{d}{k}^{-1}\frac{\log\;e}{2k}\sum\limits_{S\subseteq I_n:\;\# (S)=k}\left\{{\rm tr}
\left[\BC(S)^{-1}\bPhi(S)\right]+{\rm tr}\left[\BC(S^{\ct})^{-1}\bPhi(S^{\ct})\right]-{\rm tr}\left[\BC^{-1}
\bPhi\right]\right\}.\end{array}\eeqq
Then
\beq\label{eq:u} u(1)\geq u(2)\geq \dots \geq u(d-1)\geq u(d).\eeq
\et

\bt\label{thm:z}
{\rm{(Cf. \cite{DCT}, Theorem 35.)}}\; Under condition  \eqref{cond.MWEN}, let
\beqq \begin{array}{l} z(k)=\diy\binom{d}{k}^{-1}\frac{\alpha(\BC)}{2} \log\;\left[
\prod\limits_{S\subseteq I^{(d)}:\;\# (S)=k}\frac{({\rm {det}}\BC(S))\;({\rm {det}}\BC(S^{\ct}))}{({\rm {det}}\BC)}
\right]\\
\qquad +\diy\binom{d}{k}^{-1}\frac{\log\;e}{2}\sum\limits_{S\subseteq I^{(d)}:\;\# (S)=k}\left\{{\rm tr}
\left[\BC(S)^{-1}\bPhi(S)\right]+{\rm tr}\left[\BC(S^{\ct})^{-1}\bPhi(S^{\ct})\right]-{\rm tr}\left[\BC^{-1}
\bPhi\right]\right\}.\end{array}\eeqq
Then
\beq\label{eq:z} z(1)\geq z(2)\geq \dots \geq z(\lfloor d/2 \rfloor).\eeq
\et

\section{Weighted Hadamard-type inequalities}\label{sect:WHI}

In this section we group several results related to the weighted Hadamard inequality (WHI); cf.
\cite{SY}, Theorem 3.3. The WHI inequality asserts that for a
$d\times d$ positive definite matrix $\BC$, under condition \eqref{eq:Hada}
we have:
\beq\label{eq:Hadamard} \diy\alpha (\BC)\log \prod\limits_{i}\BC_{ii}+(\log\,e)
\sum\limits_i\BC_{ii}^{-1}\Phi_{ii}-
\alpha (\BC)\log \; {\rm{det}}\,\BC-(\log\,e){\rm{tr}}\,\BC^{-1}\bPhi \geq 0, \eeq
with equality iff $\BC$ is diagonal. Recall, 
$\alpha (\BC)=\alpha _\phi (\BC)$ and
$\bPhi=\bPhi_\BC=\Phi_{\BC,\phi}$ are as in \eqref{eq:bPhi}.

We begin with the weighted version of the strong Hadamard inequality (WSHI).  The inequality (and other bounds in this section) will involve
determinants
${\rm{det}}\,\BC (S)$ of sub-matrices $\BC (S)$ in $\BC$ where, as before, $S$ is a subset of
$I^{(d)}:=\{1,\ldots ,d\}$ of a
special type. Namely, we fix $p\in\{1,\ldots ,d-1\}$ and consider the segment $I_{p+1,d}=\{p+1,\ldots ,d\}$,
segment $I_{1,p}=\{1,\ldots ,p\}$ and unions $\{i\}\cup I_{p+1,d}$ and $I_{1,i}\cup I_{p+1,d}=I_{i+1,p}^{\ct}$
where $i\in I_{1,p}$. We deal with the related entry $C_{ii}$ in $\BC$ and  sub-matrices
$$\BC_{p+1}^d:=\BC (I_{p+1,d}), \;\BC_1^{i-1}
:=\BC (I_{1,i-1}),\;\BC (\{i\}\cup I_{p+1,d})\;\hbox{ and }\;\BC (I_{1,i}\cup I_{p+1,d})$$
and Gaussian random variables $X_i$ and vectors
$\BX_{p+1}^d:=\BX (I_{p+1,d})$, $\BX_1^{i-1}:=\BX (I_{1,i-1})$,\\ $X_i\vee\BX _{p+1}^d:=
\BX (\{i\}\cup I_{p+1,d})$ and $\BX_1^i\vee\BX_{p+1}^d:=\BX (I_{1,i}\cup I_{p+1,d})$
using symbols $x_i$, $\bx_{p+1}^d$,
$\bx_1^{i-1}$,  and $\bx_1^i\vee\bx_{p+1}^d$ for their respective values. Thus, PDFs
$$f_{\BX_{p+1}^d}(\bx_{p+1}^d)=f^{\rm{No}}_{\BC_{p+1}^d}(\bx_{p+1}^d)\hbox{ and }
f_{\BX_1^i\vee\BX_{p+1}^d}(\bx_1^i\vee\bx_{p+1}^d)=f^{\rm{No}}_{\BC (I_{1,i-1}\cup
I_{p+1,d})}(\bx_1^i\vee\bx_{p+1}^d)$$
emerge, as well as conditional PDFs $f_{X_i|\BX_{p+1}^d}(x_i|\bx_{p+1}^d)$ and
$f_{\BX_1^{i-1}|\BX_{p+1}^d}(\bx_1^{i-1}|\bx_{p+1}^d)$. Viz., $\BX_1^i\vee\BX_{p+1}^d$ and
$\bx_1^i\vee\bx_{p+1}^d$ stand for the concatenated vectors
$\begin{pmatrix}X_1\\ \vdots\\ X_i\\ X_{p+1}\\ \vdots\\ X_d\end{pmatrix}$ and
$\begin{pmatrix}x_1\\ \vdots\\ x_i\\ x_{p+1}\\ \vdots\\ x_d\end{pmatrix}$, each with $i+d-p$ entries.
As above (see \eqref{eq:bPhi}), for a given WF
$\bx\in\bbR^d\mapsto\phi (\bx )$ we consider numbers $\alpha (\BC_1^p)=\alpha_\phi (\BC_1^p)$
and matrices $\bPhi_{\BC_1^p}=\bPhi_{\BC_1^p,\BC,\phi}$:
$$\alpha (\BC_1^p)=\alpha_\phi (\BC_1^p)=\int\limits_{\bbR^d}\phi (\bx_1^d)f^{\rm{No}}_{\BC_1^p}
(\bx_1^p)\rd\bx,\;\;
\bPhi_{\BC_1^p}=\bPhi_{\BC_1^p,\BC,\phi}=\int\limits_{\bbR^d}\bx_1^p\left(\bx_1^p\right)^{\rm T}
\phi (\bx )f^{\rm{No}}_\BC(\bx )\rd\bx.$$
(In Eqns \eqref{eq:varpi} and \eqref {eq.sumCWEX} -- \eqref{eq.sumCWEY} we will use variations of
these formulas.) We also set
\beq\label{eq:phisk}\beal\bPhi_{p+1}^d \;=\;\diy\int\limits_{\bbR^{p-d}} \bx_{p+1}^d\left(\bx_{p+1}^d\right)^{\rm T} \psi (I_{p+1,d};\bx_{p+1}^d)\; f_{\BX_{p+1}^d}(\bx_{p+1}^d)\;\rd\bx_{p+1}^d,\\
\bPhi (\{i\}\cup I_{p+1,d})=\diy\int\limits_{\bbR^{p-d+1}} (x_i\vee\bx_{p+1}^d)\left(x_i\vee\bx_{p+1}^d\right)^{\rm T}\\
\quad\times\psi (\{i\}\cup I_{p+1,d};x_i\vee\bx_{p+1}^d) f_{X_i\vee\BX_{p+1}^d}(x_i\vee\bx_{p+1}^d)\rd(x_i\vee\bx_{p+1}^d),\ena\eeq
with reduced WFs $\psi (I_{p+1,d})$ and $\psi (\{i\}\cup I_{p+1,d})$ calculated as in \eqref{eq:psiS},
for $S=I_{p+1,d}$ and $S=\{i\}\cup I_{p+1,d}$.

Furthermore, we will assume in Theorem \ref{thm:WSHI} that, $\forall$ $i=1,\ldots ,p$, the reduced
WF $\psi (S)$ with $S=\{1,\ldots i,p+1,\ldots d\}=I_{i+1,p}^{\ct}$  obeys
\beq\label{eq:ExtHAss}\beal\diy\int\limits_{\bbR^{i+d-p}}\psi (I_{i+1,p}^{\ct};\bx_1^i\vee\bx_{p+1}^d)
\Big\{f_{\BX_1^i\vee\BX_{p+1}^d}(\bx_1^i\vee\bx_{p+1}^d)\\
\quad\diy -f_{\BX_{p+1}^d}(\bx_{p+1}^d)\times\Big[f^{\rm{No}}_{X_i|\BX_{p+1}^d}(x_i|\bx_{p+1}^d)
f_{\BX_1^{i-1}|\BX_{p+1}^d}(\bx_1^{i-1}|\bx_{p+1}^d)
\Big]\Big\}\rd (\bx_1^i\vee\bx_{p+1}^d)\geq 0.\ena\eeq

The `standard' SHI is
\beq\label{eq:SHi}\beal\diy\frac{{\rm{det}}\,\BC}{{\rm{det}}\,\BC_{p+1}^d}\leq\prod\limits_{1\leq i\leq p}
\frac{{\rm{det}}\,\BC(\{i\}\cup I_{p+1,d})}{{\rm{det}}\,\BC_{p+1}^d}\\
\diy\quad\hbox{ or }\;\log\,{\rm{det}}\,\BC +(p-1)\log\,{\rm{det}}\,\BC_{p+1}^d\leq \diy\sum\limits_{1\leq i\leq p}
\log\,{\rm{det}}\,\BC(\{i\}\cup I_{p+1,d}).\ena\eeq
The WE approach offers the following WSHI:

\bt\label{thm:WSHI}{\rm{(Cf. \cite{CT}, Theorem 8 or \cite{DCT}, Theorem 28.)}}
 Under condition \eqref{eq:ExtHAss}, for $1\leq  p< d$,
\beq \label{eq:WSHI}\begin{array}{l}
\diy\alpha(\BC )\log\;\left[(2\pi)^d\; {\rm{det}}\,\BC\right]+(\log\;e){\rm{tr}}\,(\BC^{-1}\bPhi )\\
\quad +(p-1)\Big\{\alpha(\BC_{p+1}^d)\diy\log\;\left[(2\pi)^{d-p}\; {\rm{det}}\,\BC_{p+1}^d\right]
+(\log\;e ) {\rm{tr}}\,[(\BC_{p+1}^d)^{-1}\bPhi_{p+1}^d]\Big\}\\
\qquad \leq\diy\sum\limits_{1\leq i\leq p}\bigg\{\alpha
(\BC (\{i\}\cup I_{p+1,d}))\log\;\left[(2\pi)^{d-p+1}{\rm{det}}\,\BC(\{i\}\cup I_{p+1,d})\right]\\
\qquad\qquad\qquad\qquad +(\log\;e){\rm{tr}}\,[\BC (\{i\}\cup I_{p+1,d})^{-1}\bPhi (\{i\}\cup I_{p+1,d})]\bigg\}.\end{array}\eeq
\et

\bp We use the same idea as in Theorem 3.3 from \cite{SY}.
Recalling (\ref{eqWCE:2}) we can write
\beqq \beal\hwphi(\BX_1^p|\BX_{p+1}^d)=\diy\frac{1}{2}\;\log\;\left[(2\pi)^d\; {\rm{det}}\,\BC\right]\;\alpha(\BC )+\frac{\log\;e}{2} {\rm{tr}}\,(\BC^{-1}\bPhi )\\
\qquad\diy -\frac{1}{2}\log\;\left[(2\pi)^{d-p}\; {\rm{det}}\,\BC_{p+1}^d\right]\;\alpha(\BC_{p+1}^d)
-\frac{\log\;e}{2} {\rm{tr}}\,[(\BC_{p+1}^d)^{-1}\bPhi_{p+1}^d],\ena\eeqq
Cf. Eqns \eqref{eq:bPhiSs},  \eqref{eq:alphaA}, \eqref{eq:alphaCk}.
Furthermore, by subadditivity of the conditional WE (see \cite{SY}, Theorem 1.4), under assumption
\eqref{eq:ExtHAss} we can write
\beq\label{eq.multi.sub} \hwphi(\BX_1^p|\BX_{p+1}^d) \leq \sum\limits_{i=1}^p h^{\rm w}_{\psi (\{i\}\cup I_{p+1,d})}(X_i|\BX_{p+1}^d).\eeq
Here for $i=1,\dots,p$, again in agreement with (\ref{eqWCE:2}),
\beqq \beal\hw_{\psi (\{i\}\cup I_{p+1,d})} (X_i|\BX_{p+1}^d)=\diy\diy\frac{1}{2}\log\left[(2\pi)^{d-p+1}{\rm{det}}\,\BC (\{i\}\cup I_{p+1,d})\right]\alpha(\BC (\{i\}\cup I_{p+1,d}))\\
\qquad\qquad\qquad\qquad\diy +\frac{\log\;e}{2} {\rm{tr}}\,\BC (\{i\}\cup I_{p+1,d})^{-1}\bPhi (\{i\}\cup I_{p+1,d})\\
\qquad\qquad\qquad - \diy\frac{1}{2}\log\;\left[(2\pi)^{d-p}\; {\rm{det}}\,\BC_{p+1}^d\right]\;\alpha(\BC_{p+1}^d)
-\frac{\log\;e}{2} {\rm{tr}}\,[(\BC_{p+1}^d)^{-1}\bPhi_{p+1}^d].\ena\eeqq
Substituting into \eqref{eq.multi.sub} yields the assertion of the theorem.
\ep

Our next result, Theorem \ref{lem:Cdr}, gives an extension of Lemma 9 from \cite{CT} (or Lemma 8 from \cite{DCT}). The latter asserts
that an individual
diagonal entry $C_{ii}$ of a $d\times d$ positive definite matrix equals the ratio of the relevant determinants, viz.,
$$C_{dd}=\frac{{\rm{det}}\,\BC}{{\rm{det}}\,\BC_1^{d-1}},\;\hbox{ or }\;\log\,C_{dd}+\log\,{\rm{det}}\,\BC_1^{d-1}-\log\,{\rm{det}}\,\BC=0.$$
Remarkably, Theorem \ref{lem:Cdr} does not require assumption \eqref{eq:ExtHAss}.

\bt\label{lem:Cdr} {\rm{(Cf. \cite{CT}, Lemma 9 or \cite{DCT}, Lemma 8.)}} The following equality holds true:
\beq\beal\diy\alpha(C_{dd})\log\big[ (2\pi)C_{dd}\big]+\alpha (\BC_1^{d-1})\log\big[ (2\pi)^{d-1} {\rm{det}}\; \BC_1^{d-1}\big]-\alpha (\BC)\log\big[ (2\pi)^d {\rm{det}}\; \BC\big] \\
\qquad\qquad\qquad \diy =(\log\; e)\;{\rm{tr}}\;\left[\BC^{-1}\bPhi\right]
-(\log\; e)\;{\rm{tr}}\;\left[\left(\BC_1^{d-1}\right)^{-1}\bPhi_1^{d-1}\right] -(\log\; e)\; C_{dd}^{-1}\Phi_{dd}.
\ena\eeq
\et

\bp Using the conditional normality of $X_d$ given $\BX_1^{d-1}$, we can write
\beqq \hwphi(X_d|\BX_1^{d-1})= \frac{\alpha(C_{dd})}{2}\log \; \big[ (2\pi)C_{dd}^2\big]+\frac{\log\; e}{2}\; C_{dd}^2\Phi_{dd}.\eeqq
On the other hand,
\beq\label{eqWCE:1} \hwphi(X_d|\BX_1^{d-1})=\hwphi(\BX_1^d)-h^{\rm w}_{\psi (I_{1,d-1})}(\BX_1^{d-1}).\eeq
and therefore 
\beq\label{eqWCE:2}\beal\diy\frac{\alpha(C_{dd})}{2}\log \; \big[ (2\pi)C_{dd}^2\big]+\frac{\log\; e}{2}\; C_{dd}^2\Phi_{dd}\\
\qquad =\diy\frac{\alpha (\BC)}{2}\log \; \big[ (2\pi)^d {\rm{det}}\; \BC\big] +\frac{\log\; e}{2}\;
{\rm{tr}}\; \BC\bPhi\\
\qquad\qquad -\diy\frac{\alpha (\BC_1^{d-1})}{2}\log \; \big[ (2\pi)^{d-1} {\rm{det}}\; \BC_1^{d-1}\big]
-\frac{\log\; e}{2}\;{\rm{tr}}\;\left[\left(\BC_1^{d-1}\right)^{-1}\bPhi_1^{(d-1)}\right].\ena\eeq
The result then follows.
\ep

The next assertion, Theorem \ref{thm:LGRCF}, extends the result of Theorem 9 from \cite{CT}
(or Theorem 29 from \cite{DCT}) that,
$\forall$ $p=1,\ldots ,d$,  $\BC\mapsto \diy\log\,\frac{{\rm{det}}\,\BC}{{\rm{det}}\,\BC_1^p}$ is a concave
 function of a positive definite $d\times d$ matrix $\BC$.
We will  write matrix $\BC$ in the block form similar to \eqref{eq:BB}:
\beq\label{eq:BBG}\BC =\begin{pmatrix}\BC_1^p&\BC_{n-p}^p\\ \BC_p^{n-p}&\BC_{p+1}^d
\end{pmatrix}.\eeq
Set $\BD\bx_{p+1}^d=\BC_{p}^{d-p}\;(\BC_{p+1}^d)^{-1} \bx_{p+1}^d$ and $\BB_1^p=\BC_1^p
-\BC_p^{d-p}\;(\BC_{p+1}^d)^{-1}\;\BC_{d-p}^p$. Consider the following inequalities
\beq\label{eq:CndCncv}\diy\int\limits_{\bbR^d}\phi (\bx )f_{\BX_1^p}(\bx_1^p)\Big[f_{\BX_{p+1}^d|
\BX_1^p}(\bx_{p+1}^d|\bx_1^p)
-f^{\rm{No}}_{\BY_{p+1}^d|\BY_1^p}(\bx_{p+1}^d|\bx_1^p)\Big]\rd\bx\geq 0\eeq
and
\beq\label{eq:CndCncv2}\begin{array}{l}\diy
\int\limits_{\bbR^d}\phi (\bx)\Big[f_\BX(\bx)-f^{\rm{No}}_{\BC}(\bx)\Big]\bigg\{\diy\log\,\left[(2\pi )^p{\rm{det}}\,\left(\BB_1^p\right)^{-1}\right]\\
\qquad\qquad\diy +(\log\,e)\left[
\left(\bx_1^p-\BD\bx_{p+1}^d\right)^{\rm T}\left(\BB_1^p\right)^{-1}\left(\bx_1^p-\BD\bx_{p+1}^d\right)\right]\bigg\}\rd\bx\leq 0.\end{array}
\eeq

\bt\label{thm:LGRCF} {\rm{(Cf. \cite{CT}, Theorem 9 or \cite{DCT}, Theorem 29.)}}
 Assume that $\BC=\lam\BC^\prime +(1-\lam )\BC^{\prime\prime}$ where  $\BC$, $\BC^\prime$
and $\BC^{\prime\prime}$ are positive definite $d\times d$ matrices and $\lam\in [0,1]$. Given a WF
$\bx\mapsto\phi (\bx)\geq 0$ and $1\leq p\leq d$, define:
\beq\label{eq:cncvrto}\beal\diy\mu (\BC)= \alpha(\BC)\log\; \left[(2\pi )^d{\rm{det}}\; \BC\right] +(\log\; e)\;{\rm{tr}}\;\left[\BC^{-1}\bPhi_\BC\right]\\
\diy\qquad\quad -\alpha(\BC_1^p)\log\; \left[(2\pi )^d{\rm{det}}\; \BC_1^p\right] -(\log\; e)\;{\rm{tr}}\;\left[\left(\BC_1^p\right)^{-1}\bPhi_{\BC_1^p}\right],\ena\eeq
and similarly with $\mu (\BC^\prime)$ and $\mu (\BC^{\prime\prime})$. 
Then
\beq\label{eq;mu} \mu (\BC )\geq\lam\mu (\BC^\prime )+(1-\lam )\mu (\BC^{\prime\prime}).
\eeq
\et

\def\BZ{\mathbf{Z}}
\def\BS{\mathbf{S}}
\def\BK{\mathbf{K}}
\def\bz{\mathbf{z}}
\def\bec{\begin{cases}}
\def\enc{\end{cases}}

\bp Again we essentially follow the method from \cite{CT} with modifications developed in \cite{SY}.  Fix two $d\times d$ positive definite
matrices $\BC^\prime$ and $\BC^{\prime\prime}$ and set $\BX^\prime\sim f^{\rm{No}}_{\BC^\prime}$, $\BX^{\prime\prime}\sim f^{\rm{No}}_{\BC^{\prime\prime}}$. Given $\lam\in [0,1]$, consider a random variable $\Theta$ taking values $\vartheta=1,2$ with probabilities
$\lam$ and $1-\lam$ independently of $(\BX^\prime,\BX^{\prime\prime})$.  Next, set
$$\BX=\begin{cases}\BX^\prime,&\hbox{when $\Theta =1$,}\\ \BX^{\prime\prime},&\hbox{when $\Theta =2$.}\end{cases}$$
Then $\BX\sim\left(\lam f^{\rm{No}}_{\BC^\prime}+(1-\lam )f^{\rm{No}}_{\BC^{\prime\prime}}\right)$ and the covariance matrix ${\rm{Cov}}\,\BX=\lam \BC^\prime+(1-\lam )\BC^{\prime\prime}=:\BC$.

With the WF $\wt\phi (\bx_1^d,\vartheta )=\phi (\bz_1^d)$, use Theorem 2.1 from \cite{SY} and Theorem \ref{Lemma 2.2} from
Section \ref{Sect:GWE} and write:
\beq\label{eq:triple}\hw_{\wt\phi}(\BX_{p+1}^d|\BX_1^p,\Theta) \leq \hwphi(\BX_{p+1}^d|\BX_1^p)\leq
 \hwphi(\BY_{p+1}^d|\BY_1^p).\eeq
 Here $\BY$ stands for the Gaussian random vector with the PDF
$f^{\rm{No}}_\BC (\bx_1^d)$.
The LHS in \eqref{eq:triple} coincides with $\lam\mu (\BC^\prime )+(1-\lam )\mu (\BC^{\prime\prime})$ and the RHS with
$\mu (\BC)$. This completes the proof.
\ep

\def\BY{\mathbf{Y}}
\def\BS{\mathbf{S}}
\def\BK{\mathbf{K}}
\def\BZ{\mathbf{Z}} 
\def\BB{\mathbf{B}}
\def\bz{\mathbf{z}}
\def\by{\mathbf{y}}
\def\Bbeta{\mathbf{\beta}}

In a particular case $p=d-1$,  the function $\BC\mapsto \diy\frac{{\rm det}\; \BC}{{\rm det}\; \BC_1^{d-1}}$
is also concave (see
\cite{CT}, Theorem 10). The weighted version of this property is encapsulated in the following result.
For a positive definite $d\times d$ matrix $\BC$ and a WF $\bx\mapsto\psi (\bx )$, set:
\beq\label{eq:varpi} \beal\diy
\varpi_\psi(\BC):=\frac{\alpha_\psi(\BC )}{2} \log \left[ (2\pi)^d{\rm{det}}\,(\BC)\right]+\frac{\log\,e}{2}{\rm{tr}}\,\left[\BC^{-1}\bPhi_{\BC,\psi}\right]\\
\qquad\quad\diy-\frac{\alpha_{\psi_1^{d-1}}(\BC_1^{d-1})}{2} \log \left[ (2\pi)^{d-1} {\rm{det}}\,(\BC_1^{d-1})\right]
-\frac{\log\,e}{2}{\rm{tr}}\,\left[\big(\BC_1^{d-1}\big)^{-1}
\bPhi_{\BC_1^{d-1},\BC,\psi}\right].\ena\eeq

\br{\rm When $\psi (\bx_1^d)\equiv 1$, the expression for $\varpi_\psi(\BC)$ in \eqref{eq:varpi}
simplifies to $\diy\log\,\frac{2\pi{\rm det}\; \BC}{{\rm det}\; \BC_1^{d-1}}$. The aforementioned concavity
 property from \cite{CT}, Theorem 10 (or from \cite{DCT}, Theorem 30), is essentially equivalent to the following subadditivity-type property:
$$\log\,\frac{2\pi{\rm det}\; (\BA+\BB)}{{\rm det}\; (\BA_1^{d-1}+\BB_1^{d-1})}\geq\log\,
\frac{2\pi{\rm det}\; \BA}{{\rm det}\; \BA_1^{d-1}} +\log\,\frac{2\pi{\rm det}\; \BB}{{\rm det}\; \BB_1^{d-1}}.$$
The WE-version of this property is more involved: see Eqns \eqref{eq:condvarpi} --
\eqref{eq:pschgam}. A crucial part is played by Lemma \ref{lem:1.3}, with $X$ represented by the random variable $Z_d\sim f^{\rm{No}}_{A_{dd}+B_{dd}}$ and $Y$ is associated with the independent Gaussian pair of vectors $(\BX_1^{d-1},\BY_1^{d-1})$ having the joint PDF
$$f_{\BX_1^{d-1},\BY_1^{d-1}}(\bx_1^{d-1},\by_1^{d-1})=f^{\rm{No}}_{\BA_1^{d-1}}(\bx_1^{d-1})f^{\rm{No}}_{\BB_1^{d-1}}(\by_1^{d-1}).$$
The random element $Z$ from Theorem \ref{lem:1.3} is represented by $\BZ_1^{d-1}$, and the map $\xi$
takes \\
$(\bx_1^{d-1},\by_1^{d-1})\mapsto \bx_1^{d-1}+\by_1^{d-1}$.}
\er

\bt {\rm{(Cf. \cite{CT}, Theorem 10 or \cite{DCT}, Theorem 30.)}}
Let $\BA$, $\BB$ be two positive definite $d\times d$  matrices and
$\BX\sim f^{\rm{No}}_\BA$, $\BY\sim f^{\rm{No}}_\BB$ be the corresponding independent Gaussian vectors,
with $\BZ:=\BX+\BY\sim f^{\rm{No}}_{\BA+\BB}$. Consider a WF\\ $(z_d,\bx_1^{d-1},\by_1^{d-1})\in
\bbR\times\bbR^{d-1}\times\bbR^{d-1}\mapsto\phi (z_d,\bx_1^{d-1},\by_1^{d-1})$ and assume the following
inequality involving conditional normal PDFs $f_{Z_d|\BX_1^{d-1},\BY_1^{d-1}}$ and
$f_{Z_d|\BZ_1^{d-1}}$:
\beq\label{eq:condvarpi}\beac\diy\int\limits_{\bbR\times\bbR^{d-1}\times\bbR^{d-1}}\phi (z_d,\bx_1^{d-1},\by_1^{d-1})
f^{\rm{No}}_{\BA_1^{d-1}}(\bx_1^{d-1})f^{\rm{No}}_{\BB_1^{d-1}}(\by_1^{d-1})\Big[f_{Z_d|\BX_1^{d-1},\BY_1^{d-1}}(z_d|\bx_1^{d-1},\by_1^{d-1})\\
\qquad\qquad -
f_{Z_d|\BZ_1^{d-1}}(z_d|\bx_1^{d-1}+\by_1^{d-1}) \Big]\rd z_d\rd\bx_1^{d-1}\by_1^{d-1}\geq 0.\ena\eeq
Then
\beq\label{eq:varpiAB} \varpi_\psi (\BA +\BB )\geq \varpi_\chi (\BA )+\varpi_\gamma (\BB).\eeq
Here
\beq\label{eq:pschgam}\beac\diy\psi (\bz_1^d)=\int\limits_{\bbR ^d}\phi (z_d-y_d,\bz_1^{d-1}-\by_1^{d-1})\frac{f^{\rm{No}}_\BA (\bz_1^d-\by_1^d)f^{\rm{No}}_\BB (\by_1^d)}{f^{\rm{No}}_{\BA+\BB}(\bz_1^d)}\rd\by_1^d,\\
\diy\chi (\bx_1^d)=\int\limits_{\bbR^d}\psi (\bx_1^d+\by_1^d)f^{\rm{No}}_\BB (\by_1^d)\rd\by_1^d,\;\;
\gamma (\bx_1^d)=\int\limits_{\bbR^d}\psi (\bx_1^d+\by_1^d)f^{\rm{No}}_\BA (\by_1^d)\rd\by_1^d.\ena\eeq
\et

\bp As in \cite{CT}, we use basic properties of Gaussian random variables. Assume $\BX\sim f^{\rm{No}}_\BA$ and $\BY\sim f^{\rm{No}}_{\BB}$ are
independent Gaussian random vectors and set $\BZ=\BX+\BY\sim f^{\rm{No}}_{\BA +\BB}$. By virtue of  \eqref{eq:GmaxCond} and Theorem \ref{lem:1.3}, we can write:
\beq\label{eq.sumCWE1}\hw_\psi(Z_d|\BZ_1^{d-1})=
\hw_\psi (\BZ)-\hw_{\psi_1^{d-1}}(\BZ_1^{d-1})=\varpi (\BA+\BB )\geq\hw_\phi (Z_d|\BX_1^{d-1},\BY_1^{d-1}).\eeq
Next, owing to independence of $\BX$ and $\BY$, the conditional WE $\hw_{\phi}(X_d+Y_d|\BX_1^{d-1},\BY_1^{d-1})$ equals the sum
\beq\label{eq.sumCWE2}\beal
\qquad\diy \int\limits_{\bbR^{d-1}}\chi_1^{d-1}(\bx_1^{d-1})f_{\BX_1^{d-1}}(\bx_1^{d-1})\bigg\{\frac{1}{2}\log\,\left(\frac{2\pi}{A^{(-1)}_{dd}}\right)\int\limits_{\bbR}\chi_d(x)f_{X_d|\BX_1^{d-1}}(x|\bx_1^{d-1})\rd x\\
\qquad\qquad\diy+\frac{\log\,e}{2}A^{(-1)}_{dd}\int\limits_{\bbR}x^2\chi_d(x)f_{X_d|\BX_1^{d-1}}(x|\bx_1^{d-1})\rd x\bigg\} \\
\qquad\diy +\int\limits_{\bbR^{d-1}}\gamma_1^{d-1}(\bx_1^{d-1})f_{\BY_1^{d-1}}(\bx_1^{d-1})\bigg\{\frac{1}{2}\log\,\left(\frac{2\pi}{B^{(-1)}_{dd}}\right)\int\limits_{\bbR}\gamma_d(x)f_{Y_d|\BY_1^{d-1}}(x|\bx_1^{d-1})\rd x\\
\qquad\qquad\diy+\frac{\log\,e}{2}B^{(-1)}_{dd}\int\limits_{\bbR}x^2\gamma_d(x)f_{Y_d|\BY_1^{d-1}}(x|\bx_1^{d-1})\rd x\bigg\}.\\
\end{array}\eeq
(The fact that $X_d$ and $Y_d$ are scalar Gaussian variables is crucial here.)

The first summand equals
\beq\label{eq.sumCWEX}\beal\diy
\frac{1}{2}\log\,\left(\frac{2\pi}{A^{(-1)}_{dd}}\right)\int\limits_{\bbR^{d-1}}\chi (\bx_1^d)f_{\BX_1^d}(\bx_1^d)\rd\bx_1^d+\frac{\log\,e}{2}A^{(-1)}_{dd}\int\limits_{\bbR^{d-1}}\chi (\bx_1^d)x_d^2f_{\BX_1^d}(\bx_1^d)\rd\bx_1^d\\
\ena\eeq
and coincides with
\beq\beal\diy \hw_\chi (X_d|\BX_1^{d-1})=
\frac{\alpha_\chi (\BA)}{2} \log \left[ (2\pi)^d{\rm{det}}\,\BA\right]+\frac{\log\,e}{2}{\rm{tr}}\,\BA^{-1}
\bPhi_{\BA,\chi}\\
\;\;\diy-\frac{\alpha_{\chi_1^{d-1}}(\BA_1^{d-1})}{2} \log \left[ (2\pi)^{d-1} {\rm{det}}\,\BA_1^{d-1}\right]
-\frac{\log\,e}{2}{\rm{tr}}\,\left[\big(\BA_1^{d-1}\big)^{-1}
\bPhi_{\BA_1^{d-1},\chi_1^{d-1}}\right]=:\varpi_\chi (\BA ).\end{array}\eeq

Similarly, the second summand coincides with
\beq \label{eq.sumCWEY} \beal\diy \hw_\gamma (Y_d|\BY_1^{d-1})=
\frac{\alpha_\gamma (\BB)}{2} \log \left[ (2\pi)^d{\rm{det}}\,\BB\right]+\frac{\log\,e}{2}{\rm{tr}}\,\BB^{-1}
\bPhi_{\BB ,\gamma}\\
\;\;\diy-\frac{\alpha_{\gamma_1^{d-1}}(\BB_1^{d-1})}{2} \log \left[ (2\pi)^{d-1} {\rm{det}}\,\BB_1^{d-1}\right]
-\frac{\log\,e}{2}{\rm{tr}}\,\left[\big(\BB_1^{d-1}\big)^{-1}
\bPsi_{\BB_1^{d-1},\gamma_1^{d-1}}\right]=:\varpi_\gamma (\BB ).\end{array}\eeq
We therefore obtain the property claimed in \eqref{eq:varpiAB}:
$\varpi_\psi (\BA +\BB )\geq\varpi_\chi (\BA)+\varpi_\gamma (\BB)$.
\ep

Finally, combining \eqref{eq:w1wd} and \eqref{eq:Hadamard}, we offer

\bt {\rm{(Cf. \cite{DCT}, Corollary 4)}} Given a
$d\times d$ positive definite matrix $\BC$, assume condition \eqref{eq:Hada}. Then
\beq \beal\diy\alpha(\BC)\log \;\left[\prod\limits_{i=1}^d \frac{2\pi({\rm {det}}\;\BC)}{{\rm{det}}\;
\BC(I_1^{i-1}\cup I_{i+1}^d)}\right]\\
\qquad\qquad\diy+\log\;e\sum\limits_{i=1}^d \left\{{\rm tr}\left[\BC^{-1}\bPhi\right]-{\rm tr}\left[\BC(I_1^{i-1}\cup
I_{i+1}^d)^{-1}\bPhi(I_1^{i-1}\cup I_{i+1}^d)\right]\right\}\\
\quad\leq\;\alpha (\BC)\log \; {\rm{det}}\,\BC-(\log\,e){\rm{tr}}\,\BC^{-1}\bPhi\;\leq\;
\diy\alpha (\BC)\log \prod\limits_{i}\BC_{ii}+(\log\,e)
\sum\limits_i\BC_{ii}^{-1}\Phi_{ii}.
\ena\eeq

\et

\vskip .5 truecm
{\bf Acknowledgement.} YS thanks the Math Department, Penn State University, for the financial support and hospitality during the academic year 2014-5. SYS thanks the CAPES PNPD-UFSCAR Foundation
for the financial support in the year 2014-5. SYS thanks the Federal University of Sao Carlos, Department of Statistics, for hospitality in 2014-5. IS has been supported by FAPESP Grant - process No 11/51845-5,
and expresses her gratitude to IMS, University of S\~{a}o Paulo and to Math Department, University of Denver, for the warm hospitality.

\end{document}